\UseRawInputEncoding
\documentclass[]{elsarticle}

\usepackage{hyperref}
\usepackage{algorithmic,algorithm}
\usepackage{color}
\usepackage{soul}
\usepackage{subfigure}

\soulregister\cite7 
\journal{Astronomy and Computing}

\begin{document}

\begin{frontmatter}

\title{Study on Outliers in the Big Stellar Spectral Dataset of the Fifth Data Release (DR5) of the Large Sky Area Multi-Object Fiber Spectroscopic Telescope (LAMOST) }%\tnoteref{mytitlenote}}


\author[1,2,3]{Yan Lu}
\author[1,2,3]{A-Li Luo\corref{lal}}
\cortext[lal]{Corresponding author}
\ead{lal@nao.cas.cn}
\author[2]{Li-Li Wang}
\author[1,2,3]{Li Qin}
\author[1]{Rui Wang}
\author[1,3]{Xiang-Lei Chen}
\author[1]{Bing Du}
\author[1,3]{Fang Zuo}
\author[1]{Wen Hou}
\author[1]{Jian-Jun Chen}
\author[4]{Yan-Ke Tang}
\author[2]{Jin-Shu Han}
 \author[1,3]{Yong-Heng Zhao}

\address[1]{CAS Key Laboratory of Optical Astronomy, National Astronomical Observatories,  Chinese Academy of Sciences, Beijing 100012, People's Republic of China}
\address[2]{College of Computer and Information Engineering \& Institute for Astronomical Science, Dezhou University, Dezhou 253023, People's Republic of China}
\address[3]{University of Chinese Academy of Sciences, Beijing 100049, People's Republic of China}
\address[4]{College of Physics and Electronic Information, Dezhou University, Dezhou 253023, People's Republic of China}

\begin{abstract}
To study the quality of stellar spectra of the Large Sky Area Multi-Object Fiber Spectroscopic Telescope (LAMOST) and the correctness of the corresponding stellar parameters derived by the LASP (LAMOST Stellar Parameter Pipeline), the outlier analysis method is applied to the archived AFGK stars in the fifth data release (DR5) of LAMOST. The outlier factor is defined in order to sort more than 3 million stellar spectra selected from the DR5 Stellar Parameter catalog. We propose an improved Local Outlier Factor (LOF) method based on Principal Component Analysis and Monte Carlo to enable the computation of the LOF rankings for randomly picked sub-samples that are computed in parallel by multiple computers, and finally to obtain the outlier ranking of each spectrum in the entire dataset. Totally 3,627 most outlier ranked spectra, around one-thousandth of all spectra, are selected and clustered into 10 groups, and the parameter density distribution of them conforms to the parameter distribution of LAMOST DR5, which suggests that in the whole parameter space the probability of bad spectra is uniformly distributed. By cross-matching the 3,627 spectra with APOGEE, we obtain 122 common ones. The published parameters calculated from LASP agree with APOGEE for the 122 spectra although there are bad pixels or bad flux calibrations in them. On the other hand, some outlier spectra show strong nebular contamination warning the corresponding parameters should be carefully used. A catalog and a spectral atlas of all the 3,627 outliers can be found at the link \url{http://paperdata.china-vo.org/LY\_paper/dr5Outlier/dr5Outlier\_resource.zip}. 
\end{abstract}

\begin{keyword}
Spectral data, Outlier Detection, Statistical analysis
\end{keyword}
\end{frontmatter}

%\linenumbers

\section {Introduction}
\label{sec:introduction}
With the increasing size and complexity of astronomical surveys, it is necessary for astronomers to take advantage of the tools developed in the fields of data science and machine learning(ML) for extracting and analysing information from ongoing and future astronomical surveys \cite{b36}. Supervised ML algorithms are usually used to detect or classify specified objects, and unsupervised ML algorithms are often useful for seeking the internal relationship between high-dimensional data and data clustering. Ball et al.(2010) provided a review of data mining and machine learning in astronomy, with a conclusion that data mining can be a powerful tool only if one carefully selects an appropriate algorithm \cite{b37}. Meusinger et al.(2012) selected a sizeable sample of unusual quasars using a method based on self-organizing maps and visual inspection of a huge number of spectra \cite{b38}. Hu et al.(2020) applied the sparse PCA algorithm for outlier detection according to the extracted variables and an optimal outlier detection model was established in \cite{b39}. Logan et al.(2020) explored the use of Random Forest and PCA as part of the pre-processing stage for feature selection and dimensionality reduction to separate stars, galaxies and QSOs \cite{b40}. 

Multi-object spectroscopy such as The Large Sky Area Multi-Object Fibre Spectroscopic Telescope (LAMOST) allows simultaneously observing hundreds of thousands of celestial objects, and the big dataset generated provides us with more opportunity to perform Galactic and extragalactic research. For example, Wu et al.(2017) obtained precise ages of 6,940 red giant branch (RGB) stars in the LAMOST-Kepler project (\cite{b16}; \cite{b18}; \cite{b17}). Ren et al.(2018) presented the DR5 catalog of white dwarf-main sequence (WDMS) binaries from LAMOST \cite{b19}. Qian et al.(2019) detected spectroscopic binary or variable candidates (SBVC) using the parameters released by LAMOST \cite{b9}. Liu et al.(2016) proposed Fuzzy Large Margin and Minimum Ball Classification Model(FLM-MBC) to accomplish the task of special celestial body exploration \cite{b13}. And Du et al.(2016) developed a method combining a bipartite ranking model with bootstrap aggregating techniques, and validated that the method is effective and less time-consuming when searching for rare spectra in a large but unlabelled data set \cite{b15}. Wang et al.(2016) proposed an outlier spectral data analysis method using line index characteristics space, and experimental results demonstrated that using line index as the characteristics value of spectrum one can quickly perform the outlier data mining for high dimensional spectral data \cite{b14}. Wu et al.(2019) proposed a method based on principal component analysis (PCA) and the density peak approach to search special stellar spectra in low-S/N (signal to noise ratio) stellar data \cite{b21}. Yang et al.(2020) designed a new lattice structure named SVM-Lattice based of SVM (Support Vector Machine) and FCL (Formal Concept Lattice) that was applied in the recognition and evaluation of rare spectra with double-peaked profiles \cite{b24}. 

Although many valuable work has been done with LAMOST spectra and parameters, the spectral specificity of these published parameters has not been summarized. The scientific goal of this paper is to analyse the selected outlier spectra and test the reliability of the released parameters corresponding to these outlier spectra, so as to verify the applicability of LASP program. 

In order to quickly find outlier spectra in millions of spectral data, we use the PCA method to first reduce the dimension of the spectra in LAMOST DR5 having released parameters. Generally, there will be information loss in the process of data dimension reduction. We prove in the method verification before the overall work that the combination of different parameters can be used to control the information loss and achieve the best statistical effect. Then we use random sampling and multi-machine parallel methods to calculate the LOF, obtain the cumulative sum of the local outlier factors of each work target, to simulate the outlier factor of each target in the overall data set. Finally, the outlier spectrum is obtained based on the outlier factor. Our method is a combination design framework formed by the fusion of existing tools, defining and designing around tasks. After dimension reduction, random sampling, outlier factor calculation in independent area, multi-machine parallel processing, and the integration of overall outlier factors, we can realize the rapid big data outlier detection process.

The paper is organized as follows: in Section 2, LAMOST DR5 spectra that are used in our work are introduced. In Section 3, we describe the main methods used for data dimension reduction and outlier detection and two parts of feasibility verification. The main work of applying the outlier analysis method to the archived AFGK stars in LAMOST DR5 is described in Section 4 and the result analysis is explained in Section 5. In Section 6, we give the conclusion.

\section {The LAMOST DR5 Observations}
\label{sect:Obs}
The Large Sky Area Multi-Object Fibre Spectroscopic Telescope (LAMOST) is a Chinese national scientific research facility operated by the National Astronomical Observatories, Chinese Academy of Sciences. LAMOST is equipped with 16 spectrographs, each of which is fed by 250 fibres. It has a specially designed Schmidt telescope with 4,000 fibres in a field of view of 20deg $^{\rm 2}$ in the sky \cite{b1}. LAMOST began a five-year regular survey in 2012 September, which includes the LAMOST ExtraGAlactic Survey (LEGAS) and the LAMOST Experiment for Galactic Understanding and Exploration (LEGUE) \cite{b2}. The main goal of the LEGUE is to obtain spectra of stars covering 9-17.8 mag (r band) in the Milky Way. As Luo et al.(2012) shows that the raw stellar spectra are first processed by the LAMOST 2D pipeline, and then LAMOST 1D pipeline performs the spectral type classification and radial velocity measurements \cite{b3}. 

The LAMOST DR5 database contains spectra obtained in the pilot survey and the first-five-years regular survey, among which there are 8,171,443 stellar spectra. Along with the spectra, six LAMOST catalogs have been published in DR5 as well including a stellar parameter catalog \cite{2019yCat.5164....0L}. The objects published in the stellar parameter catalog are A-, F-, G-, K- type stars having S/N in g band larger than 6 in dark nights, and larger than 15 in bright night. The catalog provides the basic atmospheric physical parameters like effective temperature, surface gravity and metallicity. In our work, we use spectra with published parameters in the stellar parameter catalog to search for abnormal spectra, and to study the quality of stellar spectra and the correctness of the corresponding stellar parameters derived by LASP.

\section{Main Methods and Feasibility Verification}

According to the high dimensional characteristics of spectral data, we need tools to perform dimension reduction, and Principal Component Analysis (abbreviated as PCA) is one of the applicable tools. In order to judge the outlier degree of the spectrum, we can use the Local Outlier Factor (abbreviated as LOF) method. In addition, we need to reduce the amount of data through quality control and stochastic simulation methods. So we propose an improved LOF method based on PCA and Monte Carlo method which named as ILOFPM. In this section the three basic methods and our new method are introduced before two kinds of feasibility verification.
\subsection{basic method 1: PCA}
\label{sect:PCA}
PCA method is a mathematical dimension reduction method that uses orthogonal transformation to convert a series of linearly related variables into a group of new linearly uncorrelated variables called principal components, so as to use the new variables to show the characteristics of the data in a smaller dimension. In the work of transformation the selection of new coordinate axes from the original space is closely related to the data itself. Among them, the first new axis is the direction having the largest variance in original data. The second new axis is the plane having the maximum variance orthogonal to the first axis. The third axis is the plane having the largest variance which orthogonal to the first two axes. We find that most of the variance is contained in the former $K$ axes, while the latter is almost zero. Therefore, we can retain these $K$ coordinates having the vast majority of variance, ignoring the rest of the axes.

PCA algorithm based on eigenvalue decomposition covariance matrix can be explained as following:

1.Initialization of the original matrix $X_{m*n}$.

2.Calculate the covariance matrix $Cov(X)$ of original matrix $X_{m*n}$.

3.Find out the eigenvalues and eigenvectors of covariance matrix $Cov(X)$.

4.The eigenvalues are sorted from large to small, and the largest $K$ among them are selected. Then the corresponding $K$ eigenvectors are used as row vector to form the eigenvector matrix $P$.

5.The original data is transformed into a new space constructed by $K$ eigenvectors, $Y=PX$.

Scikit-learn (sklearn for short)\footnote{ \noindent https://scikit-learn.org/stable/} is an open source machine learning library that supports supervised and unsupervised learning. It also provides various tools for model fitting, data preprocessing, model selection and evaluation, and many other utilities. In scikit-learn, PCA is implemented as a transformer object that trains data and gets $n$ component though its \underline{fit} method, and can be used on new data to project on these components. The object also provides the amount of variance explained by each of the selected components. How many principal components do we need to limit by parameter $n$ $\_$ $components$ (which means the minimum percentage threshold for the sum of the variances of the main components) in the transformer object of PCA? And how does the interpretable variance of each principal component ($explained$ $\_$ $variance$ $\_$ $ratio$) affect our work? These questions remain to be answered.

\subsection{basic method 2: LOF}

Outlier detection is generally used for anomaly detection, and it can be applied to describe the anomaly characteristics in spectral data. Traditionally the outlier detection is known as unsupervised anomaly detection. The Local Outlier Factor (LOF) method is a perfect tool for detect outlier in spectral data, and can be traced back to \cite{b6}. Here some definitions related to LOF are given as follows:

Definition1. $d(p,o)$ means the distance between two objects $p$ and $o$.

Definition2. $k$-distance means $k$-distance of an object $p$, and it is defined as the $d(p,o)$ between $p$ and an object $o$ $\in$ $D$ when the two criteria are held that (i) $d(p,o^{'})$ $\le$ $d(p,o)$ for at least $k$ objects $o^{'}$ $\in$ $D$ $\{$ $x \ne p$ $\}$ , (ii) $d(p,o^{'})$ $<$ $d(p,o)$ for at least $k-1$ objects $o^{'}$ $\in$ $D$ $\{$ $x \ne p$ $\}$.

Definition3. $N_{k}(p)$ means $k$-distance neighborhood of an object $p$, and it contains every object having a distance from $p$ not greater than the $k$-distance, including $k$-distance.

Definition4. reach-distance$_{k}(p,o)$ means the $k$th reachability distance of object $o$ to $p$ and it equals to $max$ $\{$ $k$-distance($o$),$d(p,o)$ $\}$. 

Definition5. $lrd_{k}(p)$ means local reachability density of $p$, it is defined as $lrd_{k}(p) = \frac {\left| N_{k}(p) \right| }{\sum_{ o \in N_{k}(p)} reach-distence_{k}(p,o) } $.

Definition 6. $LOF_{k}(p)$ means local outlier factor, it can be calculated by the formula: $LOF_{k}(p)=\frac {\sum_{o \in N_{k}(p)} \frac{lrd_{k}(o)}{lrd_{k}(p)}}{\left| N_{k}(p) \right|}$. This ratio usually be used to compare with a threshold that can distinguish outliers and non-outliers, and the threshold generally varies with different parameters defined in LOF calculation, with the default value 1. If the ratio is smaller than the threshold, it means that the density of $p$ is higher than that of its neighbourhood points, then $p$ is a dense point; if the ratio is higher, the density of $p$ is less than that of its neighbourhood points, and $p$ is more likely to be an abnormal point. 

For spectral data, we use the Local Outlier Factor as the outlier factor that represents how unusual and rare a spectrum is compared with other spectra. In detail, the scores obtained through LOF  module in sklearn are used as keywords for sorting our spectra. Based on experience, we use 35 as the value of $n\_neighbours$ in our work, and 10\% as the $contamination$ parameter (this parameter can help to set the proportion of outliers in the whole dataset) in our main work. Using LOF method, researchers have found some spectra with unusual continuum, some spectra having distinct characteristics like that of binary stars, emission stars, carbon stars and even some interesting spectra of unknown type. For example, Tu et al.(2009) and Tu et al.(2010) applied a LOF based outlier-detection algorithm to find out supernovae from the SDSS galaxy spectra (\cite{b10}; \cite{b11}). Wei et al.(2013) proposed a novel outlier-mining method, the Monte Carlo Local Outlier Factor (MCLOF), which was used to select outlier spectra from SDSS DR8, with a result of a total of 37,033 outlier spectra \cite{b12}.
\subsection{basic method 3: MC}
Monte Carlo (MC) method is a series algorithms that built on repeated random sampling to resolve problems, which can be divided into two categories. The problems in one category have intrinsic randomness and they can be simulated randomly. The problems in another category can be resolved by estimating characteristics of the whole sample using random sampling. We use Monte Carlo method of random sampling to estimate the LOF values of every spectrum in our dataset. In detail, we do the same operation repeatedly for all selections (we randomly select a fixed scale portion of data at each time), that approximately equivalent to operating in the whole dataset. 

\subsection{improved new method ILOFPM}
\begin{algorithm}
	\caption{ILOFPM: Improved LOF based on PCA and MC}
	\label{alg:A}
	\begin{algorithmic}
		\STATE {Input: A dataset containing above 3M spectra of high quality from LAMOST DR5 with released parameters.}
		\STATE {Output: Mean values of LOF for all spectra in the input dataset that can be used as selection criteria for outlier spectra.}
		\STATE {Begin:}
		\STATE {1.Reduce the dimension of spectra in our input dataset using PCA method.} 
		\STATE {2.The following repetitive processes were performed on 17 computers in parallel:}
		\REPEAT
		\STATE {3. For each script execution do} 
		\STATE {4. \quad For every iteration during each script execution do}
		\STATE {5. \quad \quad Random sampling to obtain a subsample.}
		\STATE {6. \quad \quad Computing the values of LOF for every spectrum in the subsample.}
		\STATE {7. \quad \quad Record the cumulative sum, subscript index and cumulative counts of LOF values related to each item in the subsample.}
		\STATE {8. \quad Reclaiming the memory in time for helping the computers return to idle state and get the temporary results.}
		\STATE {9. Get the average values of LOF for spectra that have been selected in this script execution.} 
		\UNTIL{The average number of selections for all items in the dataset reaches the threshold.}
		\STATE {End}
	\end{algorithmic}
\end{algorithm}
We propose an Improved LOF based on PCA and MC (abbreviated as ILOFPM) to obtain a kind of index to mark the degree of outlier for each spectrum in our dataset and pick out the spectra that are highly outliers. The ILOFPM method will be described in detail in the fourth section, and the special steps of the ILOFPM algorithm are shown in Algorithm 1. 

\subsection{Feasibility Verification}
\label{sect:verification}

Because of the diversity of spectral data outlier characteristics, we try to carry out a variety of samples to verify the feasibility of the work. We verify the feasibility of the work from two aspects of supervised learning and unsupervised learning.

\subsubsection{Means of Supervised Method}

First we can prove our method by means of supervised method that utilize the following information obtained by comparing the predicted labels with the true labels. The information includes:

$*$TP$:$ true positive, a spectrum belonging to the outliers is classified as an outlier.

$*$TN$:$ true negative, a spectrum not belonging to the outliers is classified as a non-outlier.

$*$FP$:$ false positive, a spectrum not belonging to the outliers is classified as an outlier. 

$*$FN$:$ false negative, a spectrum belonging to the outliers is classified as a non-outlier.

The metrics that we use in the supervised method are precision and recall:

Precision is proportion of correct positive predictions in all positive predictions. In astronomy it is common to refer to precision as purity:

Precision$=$ $\frac{TP}{TP+FP}$

Recall is proportion of truly positive predictions in all truly positive. In astronomy it is common to refer to recall as completeness:

Recall$=$ $\frac{TP}{TP+FN}$

The verification process of supervised learning method is as follows: We select 30,792 spectra with S/N in g band greater than 50 and spectral type F5 in LAMOST DR5 and name this sample set as $S1$. There are 336 spectra in $S1$ selected as outliers having different spectral anomalies manually, and the ratio of the outlier spectrum to the non-outlier spectrum is 0.011. We interpolate all spectra in set $S1$ at every 1$\mathring{A}$ in the wavelength range of 3900 - 8900$\mathring{A}$, then all of the flux are standardized in a unified way and the normalization formula is:
$ f_{i}=\frac{f_{i}}{\sum_{j=1}^{N}f_{i}^{2}} $. The numerator represents the flux value of each interpolation sampling point, and the denominator is the sum of the squares of the flux values of all sampling points in a spectrum. We set the parameter $contamination$ of LOF to 0.01, i.e. we select 1\% of the spectrum as outliers to verify the feasibility of the LOF method, and the parameter $n$$\_$$neighbors$ is set to 35 based on experience. We use Euclidean distances for near-neighbor judgment between spectral data, so we set the parameter $metric$ to `euclidean'. The spectra selected as outliers using LOF generally have the properties such as the flux values of some spectra suddenly change to 0 or abnormal values at a certain wavelength point or a range of wavelengths, or some spectra have errors or abnormal values at the splicing part of red end and blue end, which result in the spectra appearing more different from those of the ordinary spectral type F5. Through running LOF method on spectra in set $S1$, we get the recall 0.914 and the precision 1.0 which means that the LOF method can help distinguish between outlier and non-outlier. 

\subsubsection{Means of Unsupervised Method}
\begin{figure}
	\centering
	\includegraphics[width=0.4\textwidth, angle=0]{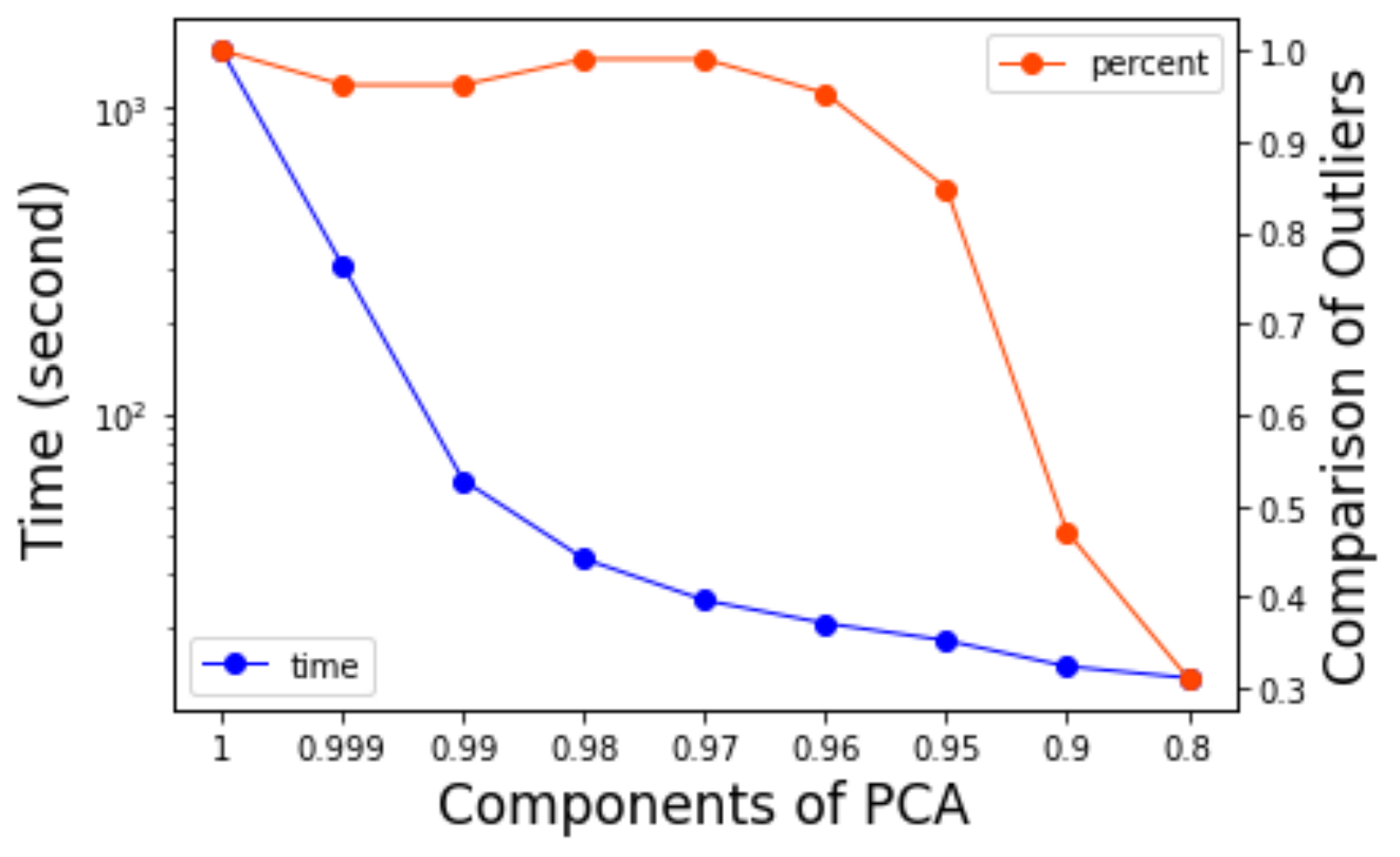}
	\caption{ The ratios of the outliers obtained through PCA and LOF to the outliers directly obtained through LOF without PCA, together with the corresponding time consumption. }
	\label{Fig1}
\end{figure}

There is such a fact that we have no outlier labels or non-outlier labels for our whole sample in hand. In addition there are many types of spectral data such as A, F, G and K in the sample, and the number and morphology of different types of spectra are diverse, so the unsupervised method is more suitable for our work. We prove it from the following steps:

The first step, we select the same spectra of stars of spectral type F5， and select other 382 spectra of quasars whose S/N in g band are greater than 10 in LAMOST DR5, which make up our sample set S2. The spectra of quasars are obviously different from that of stars of spectral type F5 in terms of spectral energy distribution, spectra features like absorption lines and emission lines, so they make up the most outlier part in S2. We experiment with S2 and compare the outliers obtained by LOF after dimension reduction of different parameters, with the main purpose to determine the parameter $n$$\_$$components$ of PCA method in the case of diversity of the outliers. We run the same interpolation and normalization processes for all spectra in set $S2$ like in $S1$. First, the LOF method is used to calculate the outlier spectra directly for $S2$ with the same parameter settings to $S1$ such as $n$$\_$$neighbours$, $contamination$ and $metric$. Second, we use PCA method to reduce the dimension of spectral data in $S2$ (by different $n$$\_$$components$ 0.999, 0.99, 0.98, 0.97, 0.96, 0.95, 0.9 and 0.8) before we calculate the outliers. We can use $n$$\_$$components$ to specify the number of dimensions that are degraded to, when $n$$\_$$components$ is an integer greater than or equal to 1. The PCA method automatically determines the number of dimensions, when $n$$\_$$components$ is a decimal float between 0 and 1. The ratio of the outliers obtained through PCA and LOF method to the outliers directly obtained through LOF is shown in $Figure~\ref{Fig1}$ attaching the corresponding time consumption in case of the parameters mentioned above. The horizontal axis represents the current parameter $n$$\_$$components$, among which, number 1 represents the result of outlier obtained directly without using dimension reduction method, and the parameter change from 0.999 to 0.8 is the minimum percentage threshold of the sum of the variances for the components. The right vertical axis shows all the results found after dimension reduction under different parameters, presented as a percentage of the first result got directly without reduction, and the values are displayed as orange dot in the figure. The left vertical axis represents the time consumption of various processes, displayed as blue dots. From $Figure~\ref{Fig1}$ we can conclude that different combinations of main components affect the number of results and it can be seen that the results obtained by 0.98 has the advantage of larger number. What's more, we get the following numbers of main components corresponding to different $n$$\_$$components$: 990, 161, 72, 44, 32, 24, 8, 2, and the corresponding time consumption are: 305, 60 ,33 ,24 , 20, 18, 14, 13 (the unit of time is seconds). So we decide to use 0.98 as the value of $n$$\_$$components$ for PCA in the following work.
	
\begin{figure}
	\centering
	\includegraphics[width=0.35\textwidth, angle=0]{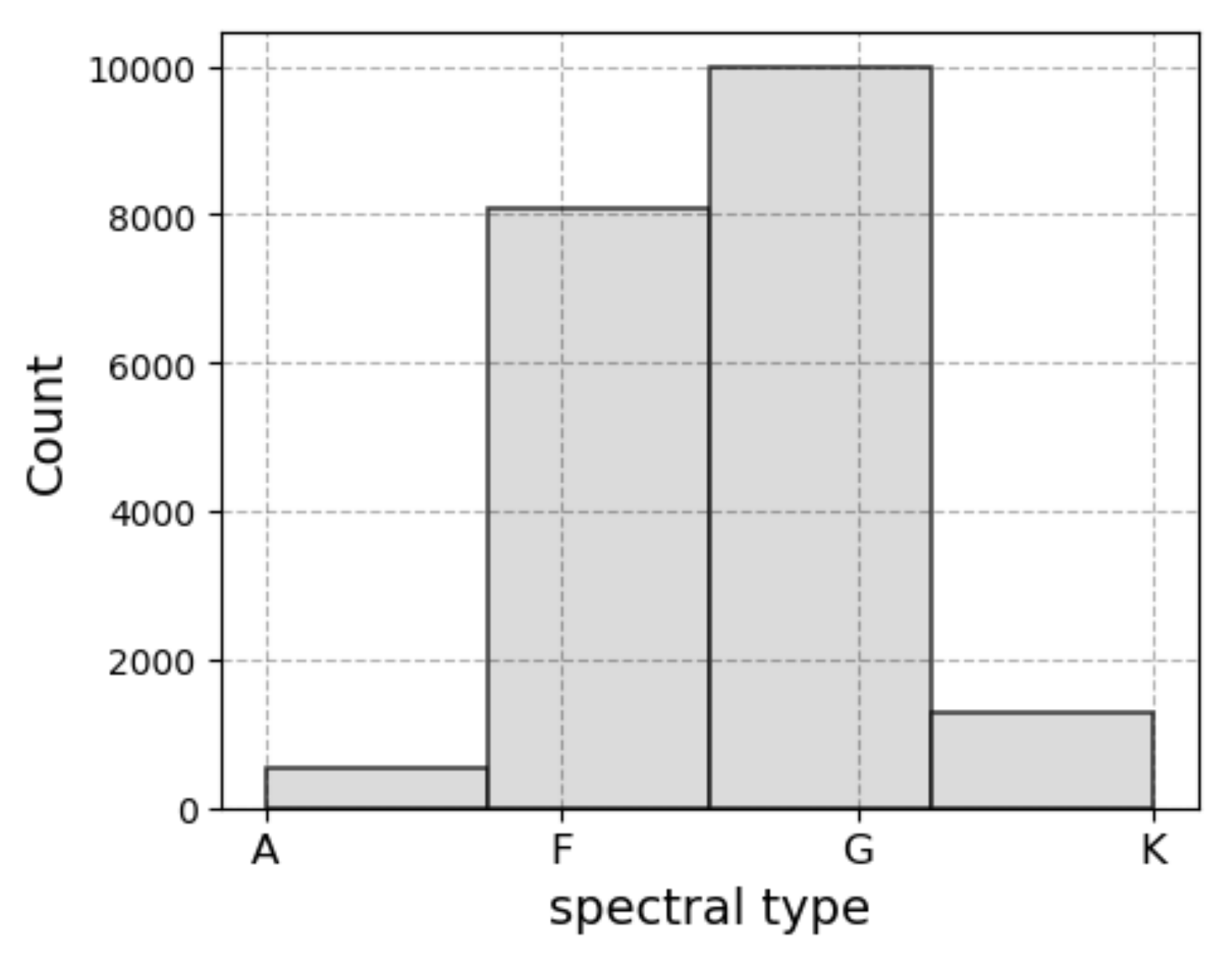}
	\caption{ The distribution of spetral type for all spectra in dataset $S3$ which be used to train the PCA model in order to get the pricipal components and the corresponding variance contribution rates. }
	\label{Fig2}
\end{figure}

\begin{figure}
	\centering
	\includegraphics[width=0.4\textwidth, angle=0]{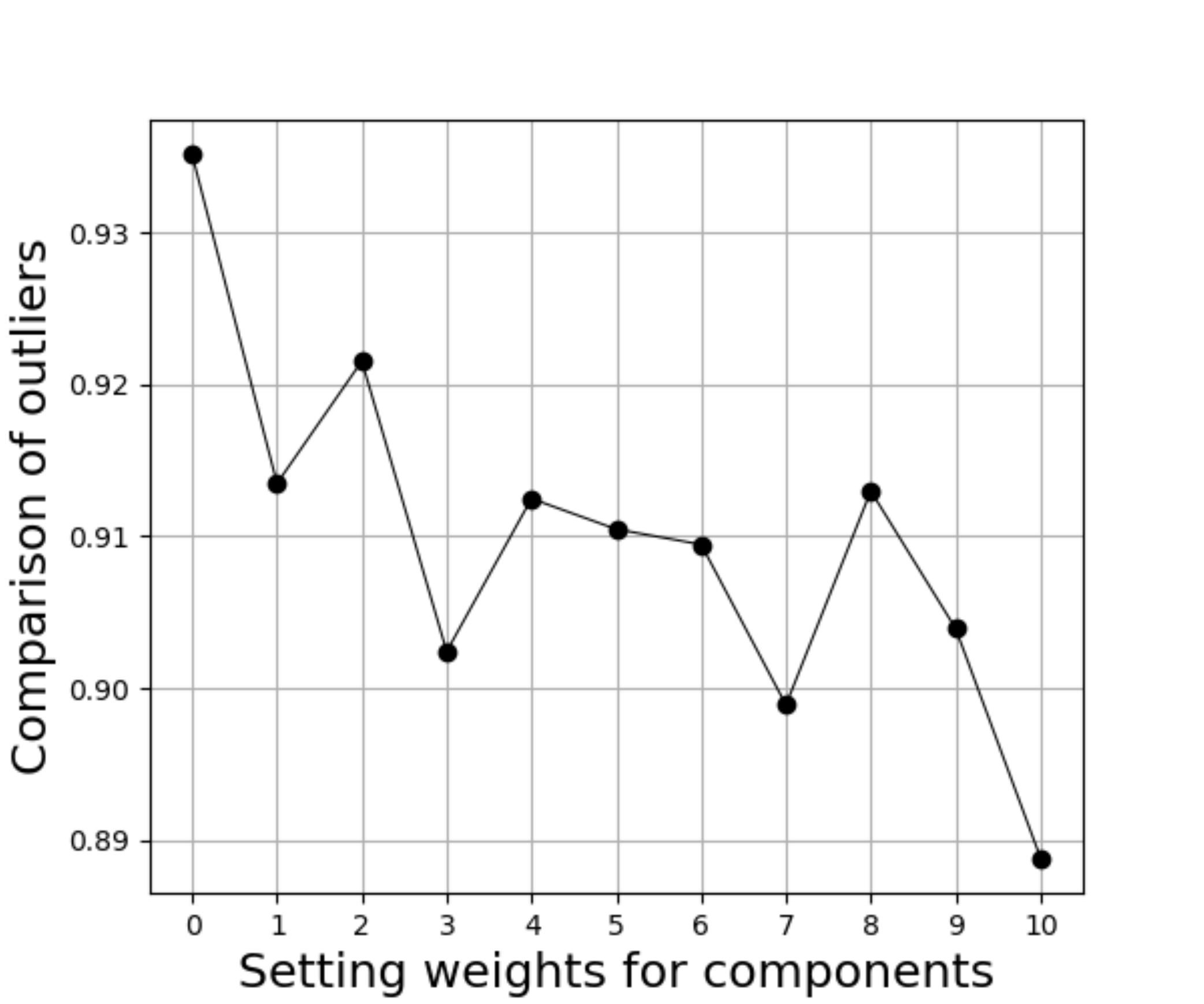}
	\caption{ 11 common parts are obtained by comparing the outliers calculated by 11 variance weighted LOF method with the outliers directly calculated by LOF without weight. The figure shows 11 ratios that are obtained by taking the number of the unweighted outliers as the denominator and the number of the 11 common parts as the numerator. }
	\label{Fig3}
\end{figure}

Then the second step, we will verify the influence of each principal components of PCA on the outlier result. In the spectra of LAMOST DR5 which have released parameter such as temperature, gravity and metallicity, we get a prepared sample dataset $PD$ (see $ \ref{sect:prepare} $ for details) and we use the three selection criteria which are explained in $ \ref{sect:prepare} $. And from the sample dataset $PD$ we select the top 20,000 spectra as a new spectra dataset named $S3$ to verify the influence of different principal components of PCA on outlier detection. The distribution of spectral type for $S3$ is shown in $Figure~\ref{Fig2}$. For dataset $S3$, we carry out interpolation and standardization operations on each spectrum in the same way we do for $S1$ and $S2$. By setting the parameter $n$$\_$$components$ to 0.98, we run the  \underline{fit} function on dataset $S3$ to get the PCA model together with 12 principal components (see \ref{sect:pca} for details) and the corresponding variance contribution rate. In order to check the influence of each principal component on the outlier detection, we use the variance contribution rates as the weights to restrict the principal components and the outlier indices are calculated through LOF method according to different weight setting strategies. Firstly, the weight of the first principal component is set to 0, and the weights of other positions are set to corresponding variance contribution rates, then the outlier spectra are obtained through LOF method for this operation. Secondly, the weights of the first and the second principal components are set to 0, and the other weights of other positions are set to corresponding variance contribution rates, then we get the outliers for the second operation. Such operations go on. Finally, only the last weight is set to corresponding variance contribution rate and all the others are set to 0, and we get the outliers for the last operation. So 11 different variance weighted LOF calculations are performed totally, and 11 common parts are obtained by comparing the outliers calculated by 11 variance weighted LOF method with the outliers directly calculated by LOF without weight. Then 11 ratios are obtained by taking the number of the unweighted outliers as the denominator and the number of the 11 common parts as the numerator, as shown in Figure~\ref{Fig3}. On the horizontal axis, 0 to 10 indicate the weights setting strategies of the first to the eleventh principal components. For example, 0 indicates that the weight of the first principal component is set to 0. Number 1 indicates that the weights of the first and the second principal components are set to 0 ...... and 10 represents the weights from the first to the eleventh principal components are all set to 0 (only the last component position is left, which getting the corresponding variance contribution rate as the weight). Values on the vertical axis show the ratios of common parts of the outliers obtained under weight setting strategies and the outliers obtained without weight setting. It can be seen from the figure that most of the ratios of common parts are more than 90$\%$, and when only one principal component is left, the ratio drops to about 89$\%$. It shows that the influence of PCA method on the original data in the direction of the obtained principal components is gradually reduced, but the importance of these 12 principal components in our spectra data processing can not be ignored. So we will use all the 12 principal components in our main work.
	
Through the verification of dataset $S1$, $S2$ and $S3$, we can get the conclusion that it is feasible to use PCA method to reduce the dimension of spectral data and obtain the outlier spectra using LOF method on the dimension reduced spectral data.
\section {Application of our method in LAMOST DR5}
\label{sect:data}
In this section, we will introduce our main work of applying the outlier analysis method to the archived AFGK stars in LAMOST DR5, in order to study the quality of LAMOST stellar spectra and the correctness of the corresponding stellar parameters derived by the LASP (these parameters include: $T$ $_{eff}$, log $g$, $[$Fe$/$H$]$, radial velocities (RV)) (\cite{b5}; \cite{b4}). The outlier factor is defined in order to rank more than 3 million stellar spectra selecting from the DR5 Stellar Parameter catalog. Our method is computed in parallel by multiple computers, and we obtain the ranking of each spectrum in the entire dataset. Then the top 10\% of the spectra with the high outlier are used to decide a LOF cut and we get an outlier spectra dataset for further research. 

\subsection {Data sample and spectral processing}
\label{sect:spectral processing}
In this study, we download spectra in LAMOST DR5 which have released parameters such as temperature, gravity and metallicity, and our whole dataset has 5,349,401 independent spectra noted as dataset $AD$. And the spectra in dataset $AD$ belongs to the A,F,G and K type star catalog of LAMOST DR5 as mentioned in $\ref{sect:Obs}$. The spectra of LAMOST have a wavelength range of 3650-9000$\mathring{A}$ and a resolution power $R$ $\sim$ 1800. According to LAMOST DR5 release document, the value of flag ``ormask'' equalling 0 represents the corresponding sampling point in a spectrum having good quality. And we only select spectra that there are more than $\frac{2}{3} $ good quality sampling points of the total in each spectrum. After cutting the spectral data in dataset $AD$ with the criteria we obtain 3,134,236 spectra of high quality with released parameters from LAMOST DR5, that is the spectral dataset for our main work. Next we remove the sampling points not belong in the wavelength range between 3900$\mathring{A}$ and 8900$\mathring{A}$ in each spectrum, and all the spectra are interpolated to 3900-8900$\mathring{A}$ to obtain the corresponding flux with step of 1$\mathring{A}$. And then all of the flux are standardized in a unified way and the normalization formula used is:
$ f_{i}=\frac{f_{i}}{\sum_{j=1}^{N}f_{i}^{2}} $ as $\ref{sect:verification}$.
We have verified the feasibility of PCA and LOF method in the above part $\ref{sect:verification}$. So we use the PCA method to reduce the dimension of spectral data from LAMOST and the outlier indices are calculated by LOF method for outlier detection, with the aim to analyse the particularity of the outlier spectra and the reliability of their parameters more comprehensively.
\subsection {Data Dimensional Reduction}
\label{sect:data reduction}
As explained in $\ref{sect:PCA}$, scikit-learn provides various tools supporting supervised and unsupervised learning, and PCA\footnote{ \noindent https://scikit-learn.org/stable/modules/decomposition.html} is one effective data dimension reduction tool. For our dataset in LAMOST DR5, we first select 26,197 spectra from the whole dataset for PCA tool, second the \underline{fit} method is run to get the PCA model and $n$ principal components, then we apply the obtained model to the overall dataset, that is, all the spectra are projected to the principal components, and the dimensionality reduction is completed.
\subsubsection{Prepare for PCA}
\label{sect:prepare}

\begin{figure}
	\centering
	\subfigure
	{	\begin{minipage}{5.5cm}
			\flushleft
			\includegraphics[scale=0.2]{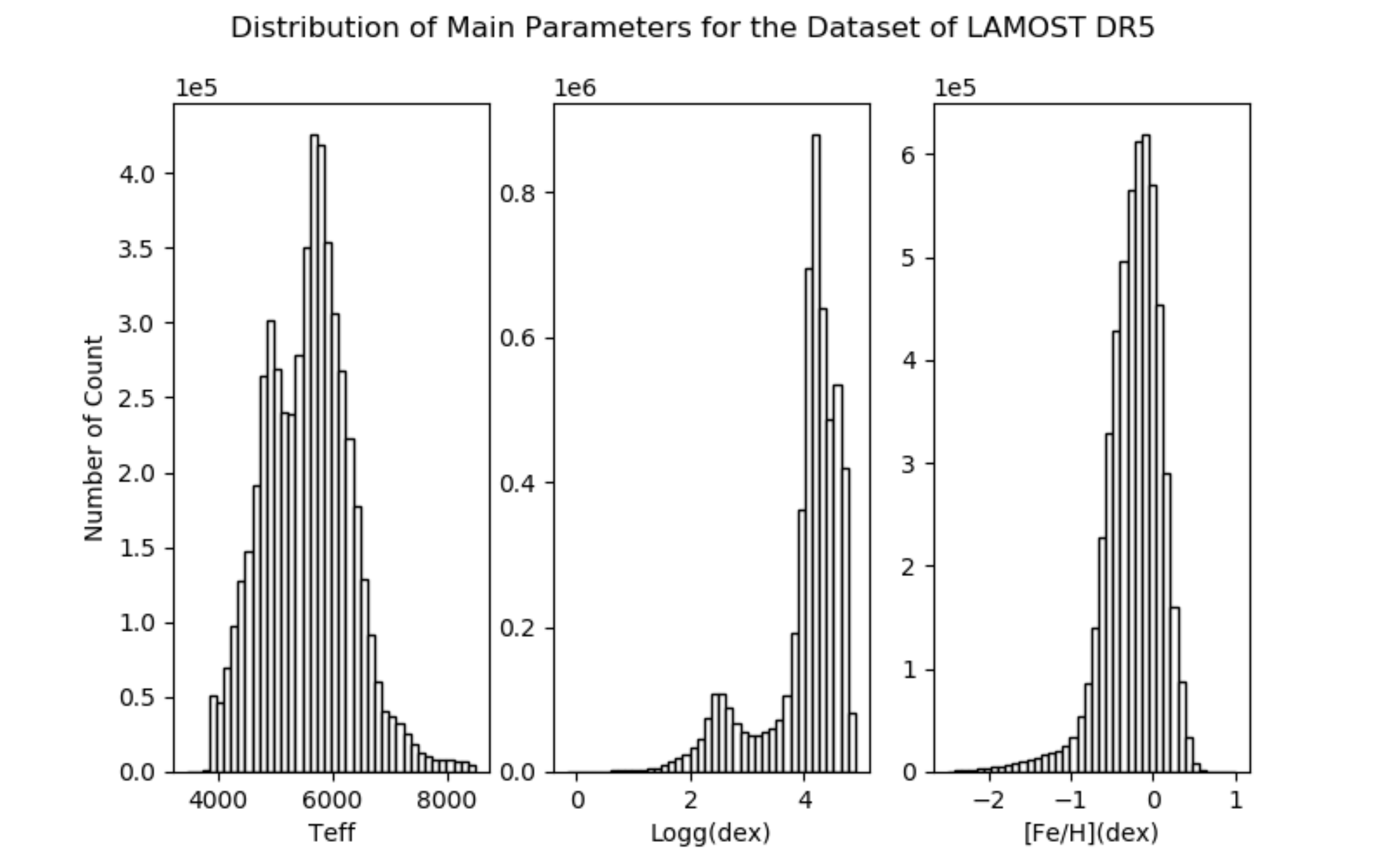}
		\end{minipage}
	}
	\subfigure
	{	\begin{minipage}{5.5cm}
			\flushright
			\includegraphics[scale=0.2]{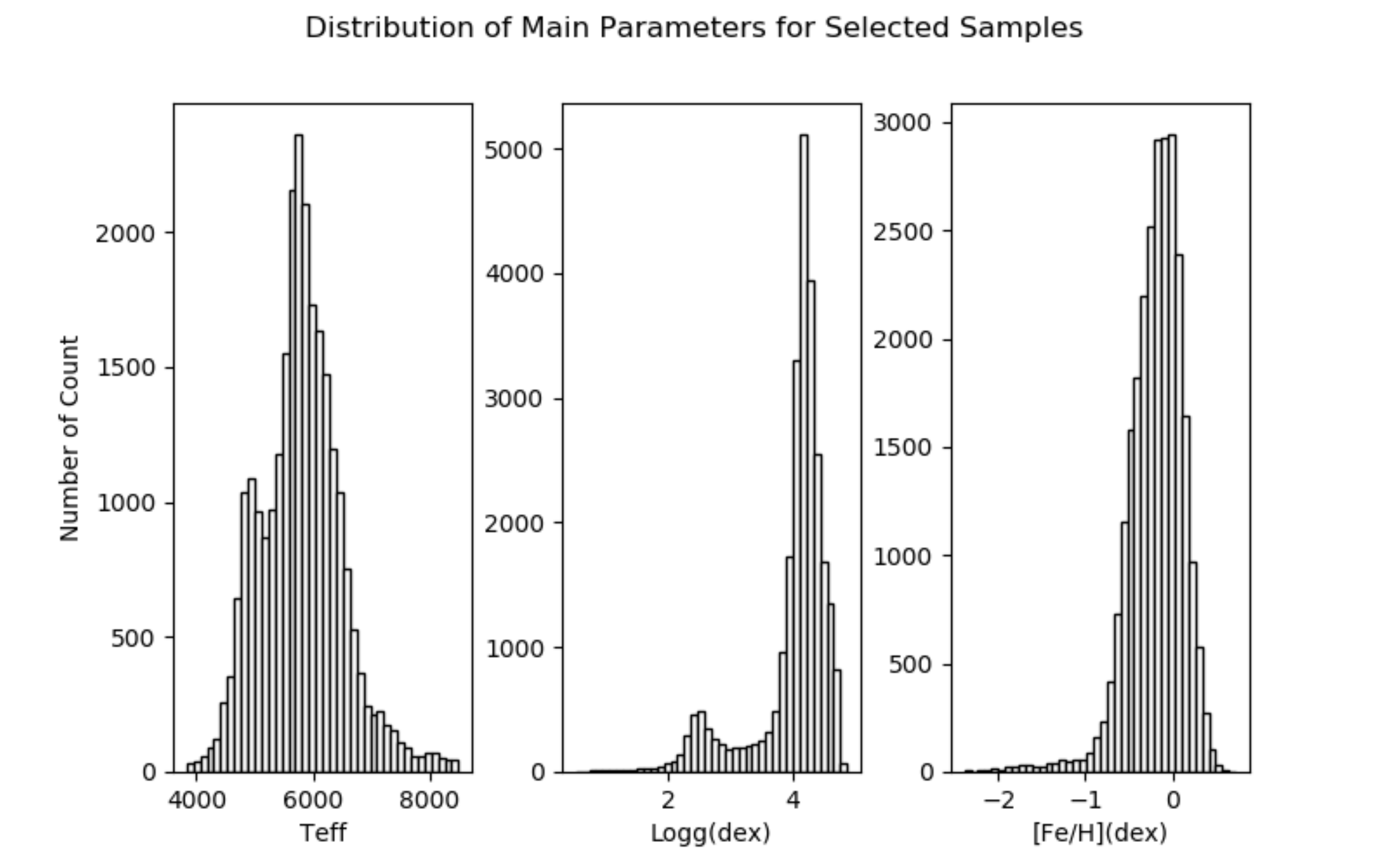}
		\end{minipage}
	}
	\caption{The left figure is the distribution of three main parameters, temperature, log gravity and metallicity of the whole dataset in LAMOST DR5 which has 5,349,401 independent spectra. And the right one is distribution of three main parameters of the randomly selected sample dataset $PD$ containing 26,197 spectra used for training PCA, and this figure shows similar distribution as the whole dataset of the left figure.}
	\label{Fig4}
\end{figure}
\begin{figure}
	\centering
	\includegraphics[width=0.4\textwidth, angle=0]{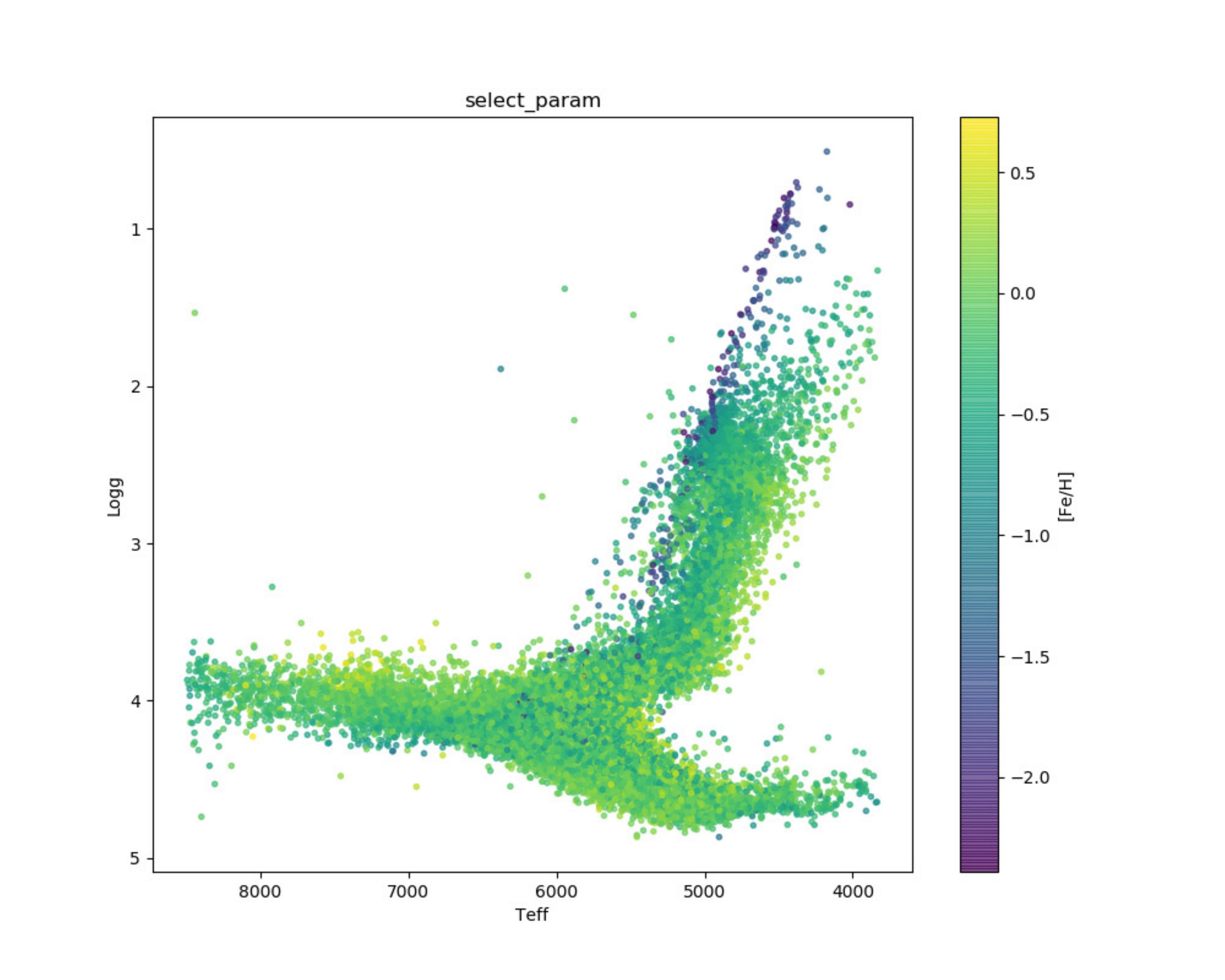}
	\caption{ HRD for the selected sample dataset $PD$ shows the distribution of log gravity vs temperature of the selected spectra. We can see from it that the selected sample dataset includes spectra of stars of various evolution stages.}
	\label{Fig5}
\end{figure}

In the part of preparing for PCA, we tentatively select 30,000 spectra from our dataset $AD$ randomly, and carry on the selection criteria for these spectra including three steps as following:
	\begin{enumerate}
		\item[(1)] The mean S/N of each spectrum must be not less than 30.
		\item[(2)] The selected spectra must have no negative flux value. 
		\item[(3)] There are more than $\frac{2}{3} $ good quality sampling points of the total in each spectrum.
	\end{enumerate}
After screening through the three selection criteria we get 26,197 spectra composing a sample dataset which named as $PD$. After normalization of the flux we get prepared for data reduction process of PCA method. The distributions of three main parameters, temperature, log gravity and metallicity of the whole dataset of LAMOST DR5 and of the selected sample $PD$ are explained in $Figure~\ref{Fig4}$. The two figures show that they have similar distribution on the three parameters, which proves that the data sample we selected is an effective representative of the overall data sample. $Figure~\ref{Fig5}$ shows the distribution of log gravity vs temperature (called Hertzsprung-Russell diagram, short for HRD) of the selected sample, and the distribution locations of our selected spectra in the HR diagram shows that the sample includes the spectra of stars of various evolution stages, which proves the validity of the sample again.
\subsubsection {Process of PCA}
\label{sect:pca}
\begin{figure}
	\centering
	\includegraphics[width=0.4\textwidth, angle=0]{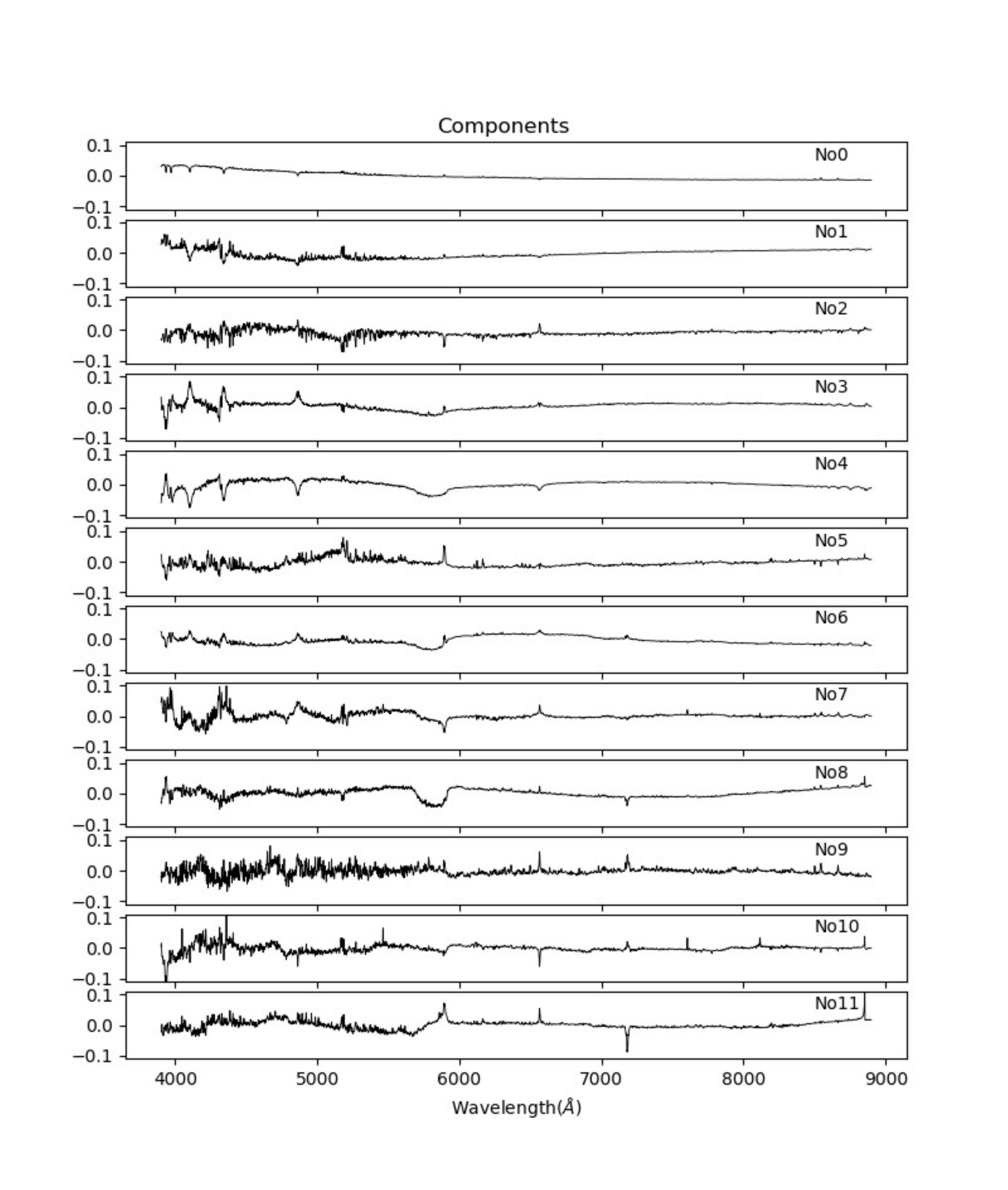}
	\caption{ The 12 principal components obtained after training the 26,197 spectra in the selected sample dataset to get PCA model, all of which will be used to reduce the dimension of the spectra in  LAMOST DR5. }
	\label{Fig6}
\end{figure}

In the step for our spectral dimension reduction, the prepared spectra got in the first step are used to train PCA model. As we explain in \ref{sect:verification} that both the time consumption and the ratio of outliers have good statistical performance when parameter is set to 0.98. So by setting the $n$$\_$$components$ to 0.98, we run the \underline{fit} method to get the PCA model and in this process we get 12 principal components. This number is smaller than the number of principal components obtained in the verification section of the first part of $\ref{sect:verification}$, because that the data objects used to train the PCA model are different in the two parts. Also in the second part of \ref{sect:verification} we explain that the importance of all these 12 principal components in our spectra data processing can not be ignored. Thus we use the 12 principal components to retain the characteristics that best reflect the individual differences of the spectra, thus we reduce the dimension of our data from several thousand to 12. The 12 principal components obtained are shown in Figure~\ref{Fig6}. 
As mentioned above, we carry out quality control process by cutting the spectral data in dataset $AD$ commended in \ref{sect:spectral processing} and we get 3,134,236 normalized spectra with released parameters from LAMOST DR5. Next we apply the obtained PCA model to these spectra and transform them into 12 dimensions. Thus we get our dimension reduced dataset named as set $RD$ which contains 3,134,236 items. In the following part, we will pick out outlier spectra from dataset $RD$.

\subsection {Outlier selecting using LOF}

In this part we will select outlier spectra with abnormal characteristic from the dataset $RD$ using LOF method.

\subsubsection{Local Outlier Factor}
\label{sect:LOF}

As we explain in \ref{sect:verification},  the LOF method can be applied to describe the abnormally of a spectrum in spectral data and we prove that it is credible to use LOF method in spectra for outlier detection. When using LOF method, a threshold is critical for generating the outlier factor based on a certain distance definition. And the threshold be controlled by the $contamination$ parameter, through which we can set the proportion of outliers in our whole dataset. In detail, we use Local Outlier Factor method (or LOF) from sklearn \footnote{ \noindent https://scikit-learn.org/stable/modules/generated/sklearn.neighbors.\\LocalOutlierFactor.html} to pick out outliers in our dataset. For preliminary outlier analysis, we try to pick out the top 10 percent with the highest outlier score in dataset $RD$, that means we set the $contamination$  a value 0.1.

Next we will introduce the process of ILOFPM (Improved LOF based on PCA and MC ).

\subsubsection {Process of ILOFPM}
\label{sect:ILOFPM}
\begin{figure}
	\centering
	\includegraphics[width=0.35\textwidth, angle=0]{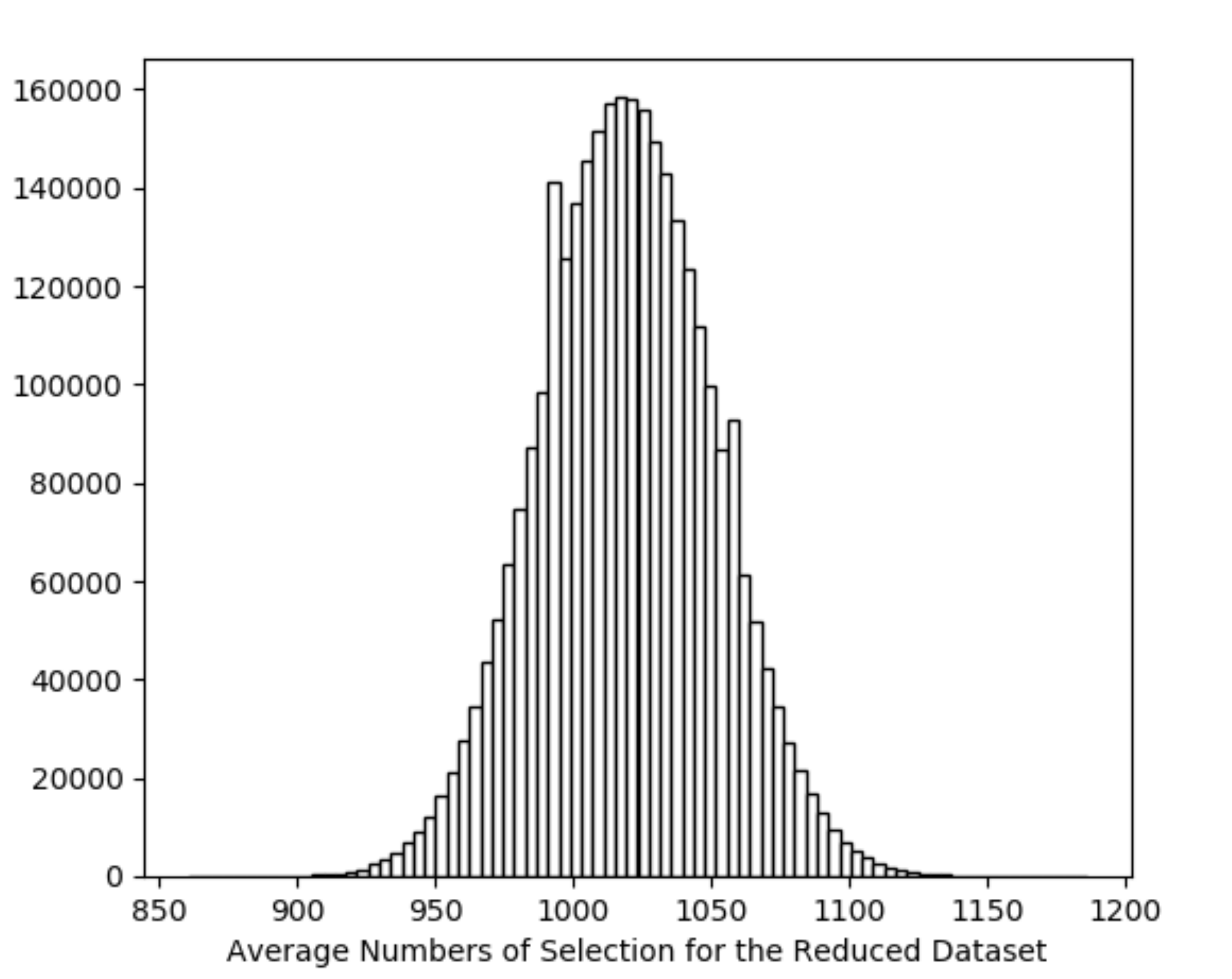}
	\caption{ Distribution of the average number of selection for the reduced dataset and all the selection numbers are greater than 900 and less than 1,140. }
	\label{Fig7}
\end{figure}

\begin{figure}
	\centering
	\includegraphics[width=0.35\textwidth, angle=0]{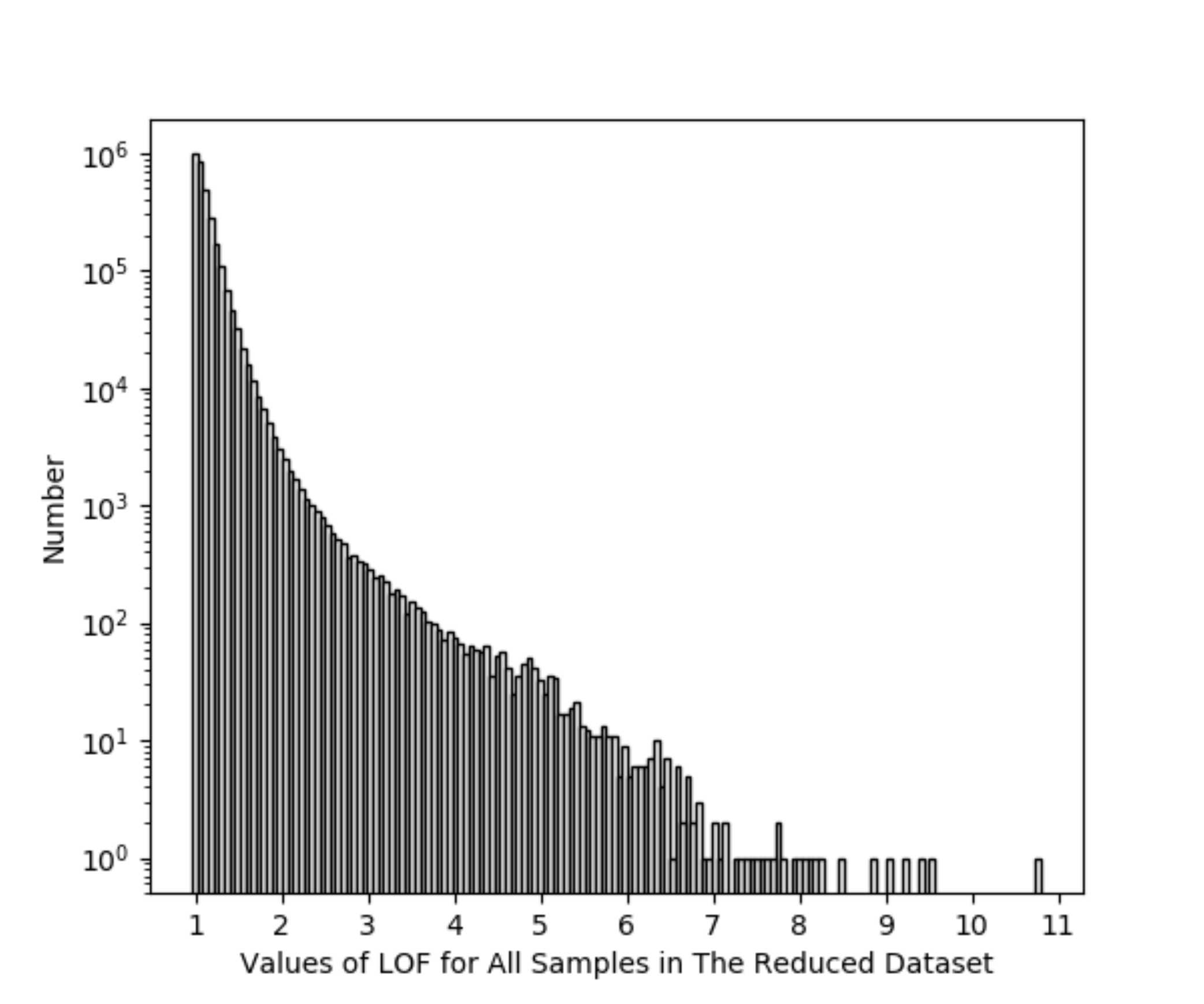}
	\caption{ All the LOF values for the reduced dataset resulted from 17 machine after process of ILOFMC. }
	\label{Fig8}
\end{figure}

In \ref{sect:data reduction} we have got our reduced dataset $RD$ through quality controlling cut and PCA execution. Another method, Monte Carlo, is used to achieve multiple random sampling, simulating the case of overall data participating in operation. For our dimension reduced dataset $RD$ from LAMOST DR5, we use random sampling to obtain a sub sample with size 90,000 each time. We calculate the values of LOF in one sub sample, and store the values for every sample along with its index belonging to the total reduced dataset $RD$. The third item we need to store is the count number for every index. During the process of repeating these operations on the same computer, if one index appears first time, we store the count value 1 and the result of outlier factor for it, and when one index appears twice or more time, we add the corresponding count numbers and get the sum value of outlier factor for each index. After 10 repetitions of execution, we get temporary results to save and finish the program to make the machine return to idle state. We script the 10 repetitions and execute them repeatedly on one machine each time. The reason to end a program after running a script is to enable the computer to reclaim memory in time and return to idle state, so it can maintain the same high efficiency when running the script program again. Even though the method of repeatedly executing the LOF to save temporary results and ending the program in time can make the program execute smoothly in script form, the time consumption on one machine is too much and can not meet our work needs. So we come up with another way to help speed up, that is to use multiple machines to take out the script process at the same time, thus we can achieve our goal faster. We use another 16 machines copying script files to them, and execute scripts 200 times on each machine manually, so we make 17 machines perform random sampling of LOF process at the same time. When the script execution process on the other 16 machines is over, we terminate the script execution on the first machine. The script on the first machine has run 350 times successfully and gets the corresponding intermediate result of 350 times. Thus the average number of selection for every one in dataset $RD$ can be derived from formula $(200*10*16+350*10)*90000/3134236$, and the result is about 1,019. We obtain the mean outlier factor of every spectra in set $RD$ using all results from the 17 machines. All the above processes are called Improved LOF method based on PCA and MC, or ILOFPM briefly. The distribution of the average selection numbers for the reduced dataset $RD$ is shown in Figure~\ref{Fig7}, and it can be found that the selection numbers are greater than 900 and less than 1,140. Figure~\ref{Fig8} shows all the outlier factor values for our reduced dataset $RD$ resulted from 17 machine after the process of ILOFPM. 

As we point out in \ref{sect:LOF} that we try to pick out the top 10 percent of all dataset and we assign the $contamination$ parameter a value 0.1, and after cutting with that parameter we get the threshold at about 1.289. Those having outlier factor values not less than the threshold are defined as outliers dataset and the rest are ordinary dataset. We get 313,424 outliers at all and give them the name $OD$. Next, we will go on the study of outlier spectra in set $OD$ in order to analyze the particularity of them.
\section {Result analysis}
\label{sect:analysis}

In this part, we first make an overall evaluation of those spectra defined as outliers, including comparison of the basic atmospheric physical parameters, S/N and RV released from LAMOST DR5 between outlier and other ordinary spectra. The mean values of S/N of outliers is lower than the ordinary spectra, and the standard deviation of outliers’ RV is a little larger than that of the ordinary. Then, according to the cumulative function graph of the sorted values of LOF, we intercept the highest part of outliers which have the highest sorted value of LOF. Using these outliers we compare the parameters of common stars from crossing match with APOGEE. The result show that the stellar parameters of them are mostly corrected according to external comparison even the spectra have lots of bad pixels or bad flux calibrations, which means the LASP has good adaptability. Then, we divide these most outlier spectra into 10 clusters according to different spectral specificity. And the distributions of the parameters of the 10 groups in the parameter grid of LAMOST DR5 empirical library are used to analyse the properties of the outlier spectra.

\subsection {Comparison between outliers and the ordinary}

\begin{figure}
	\centering
	\includegraphics[width=0.45\textwidth, angle=0]{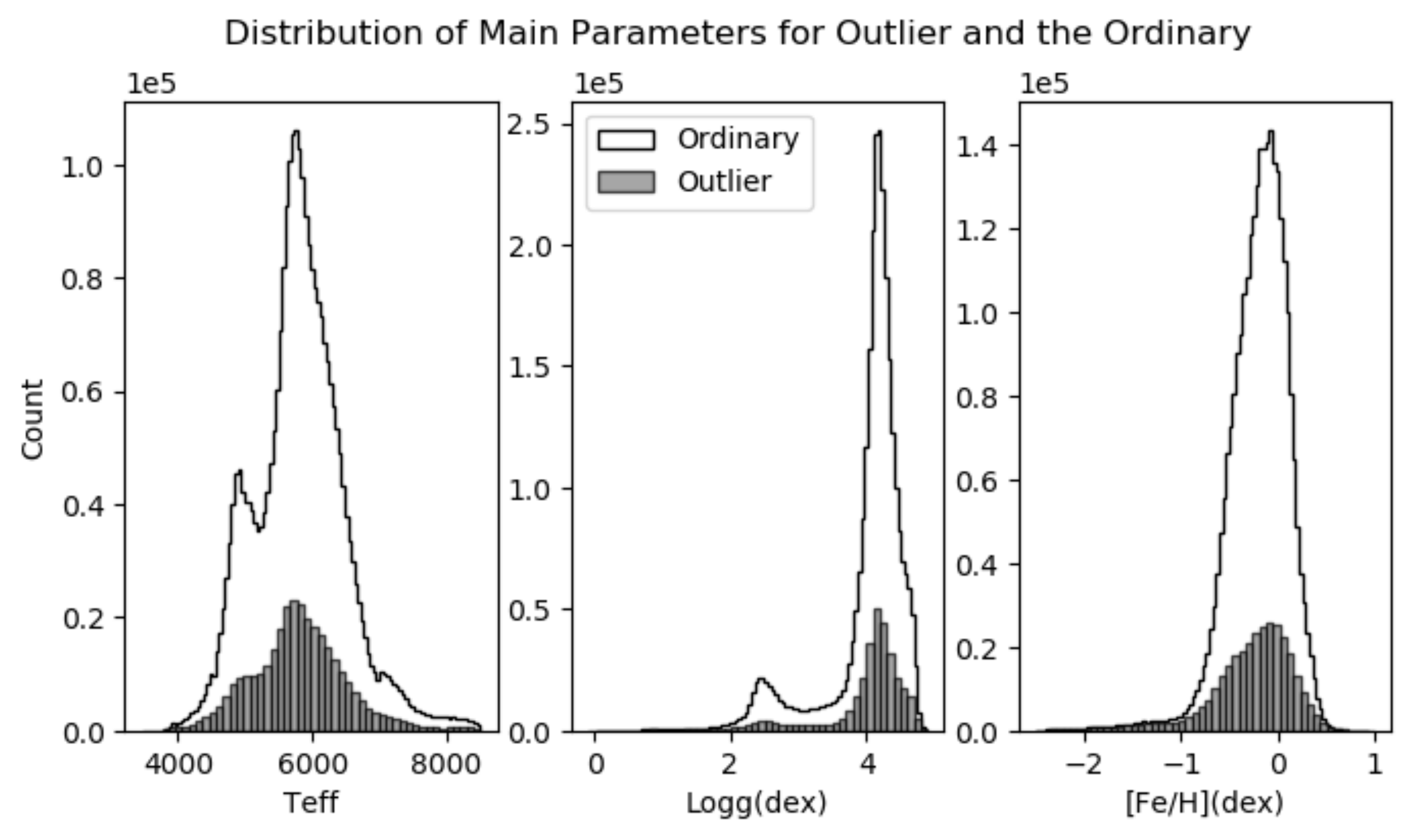}
	\caption{ Comparison of basic atmospheric physical parameters of Outliers (313,424 spectra) and the Ordinary (2,820,812 spectra). Those having values of LOF not less than the threshold are defined as outliers and the rest are ordinary.  }
	\label{Fig9}
\end{figure}

\begin{figure}
	\centering
	\includegraphics[width=0.4\textwidth, angle=0]{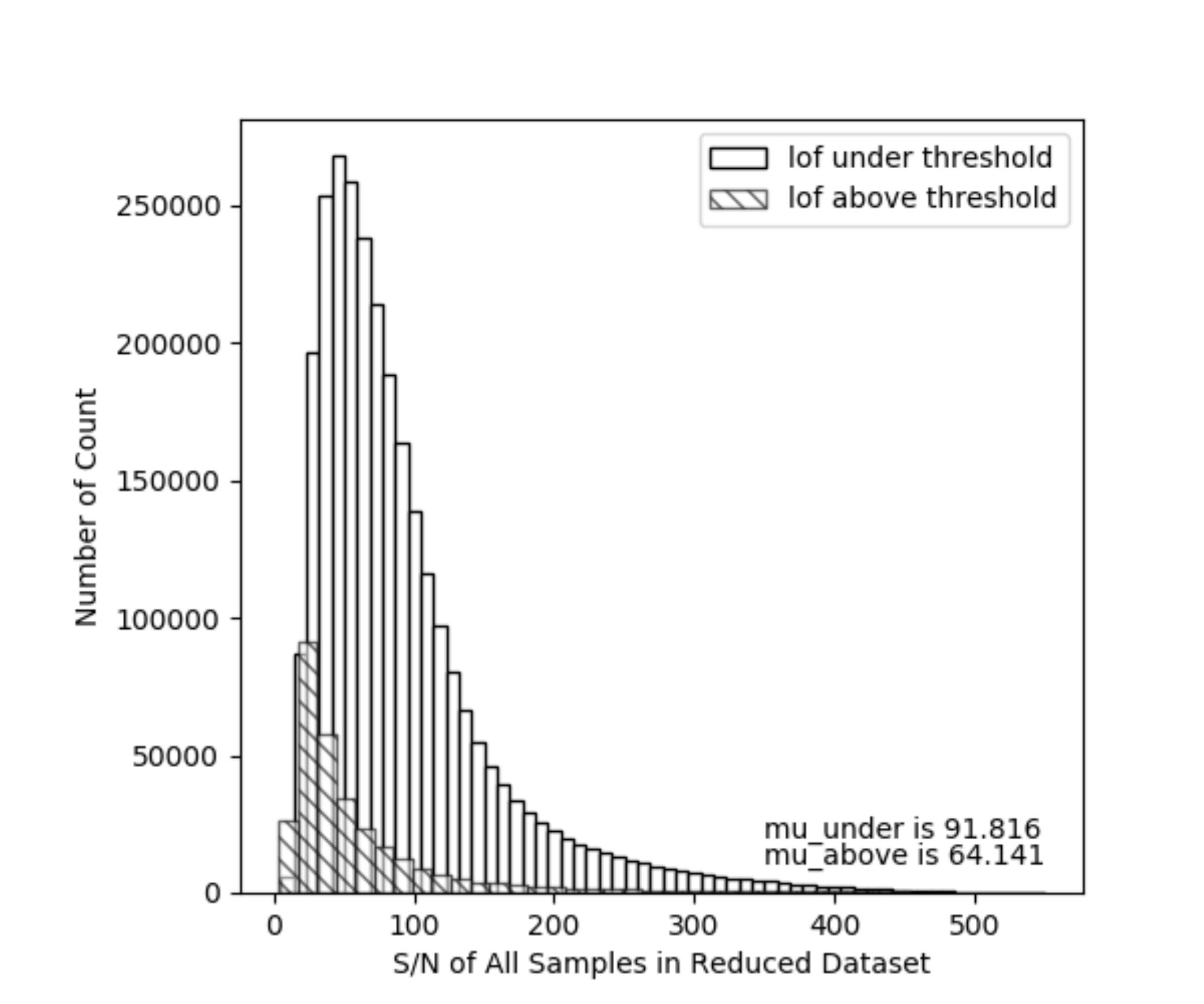}
	\caption{ Comparison of S/N of Outliers (313,424 spectra) and the Ordinary (2,820,812 spectra), and the numbers of respective bins are 40 and 60. And the mean values of S/N for outliers and the rest are displayed at the bottom right of the figure. }
	\label{Fig10}
\end{figure}

\begin{figure}
	\centering
	\includegraphics[width=0.4\textwidth, angle=0]{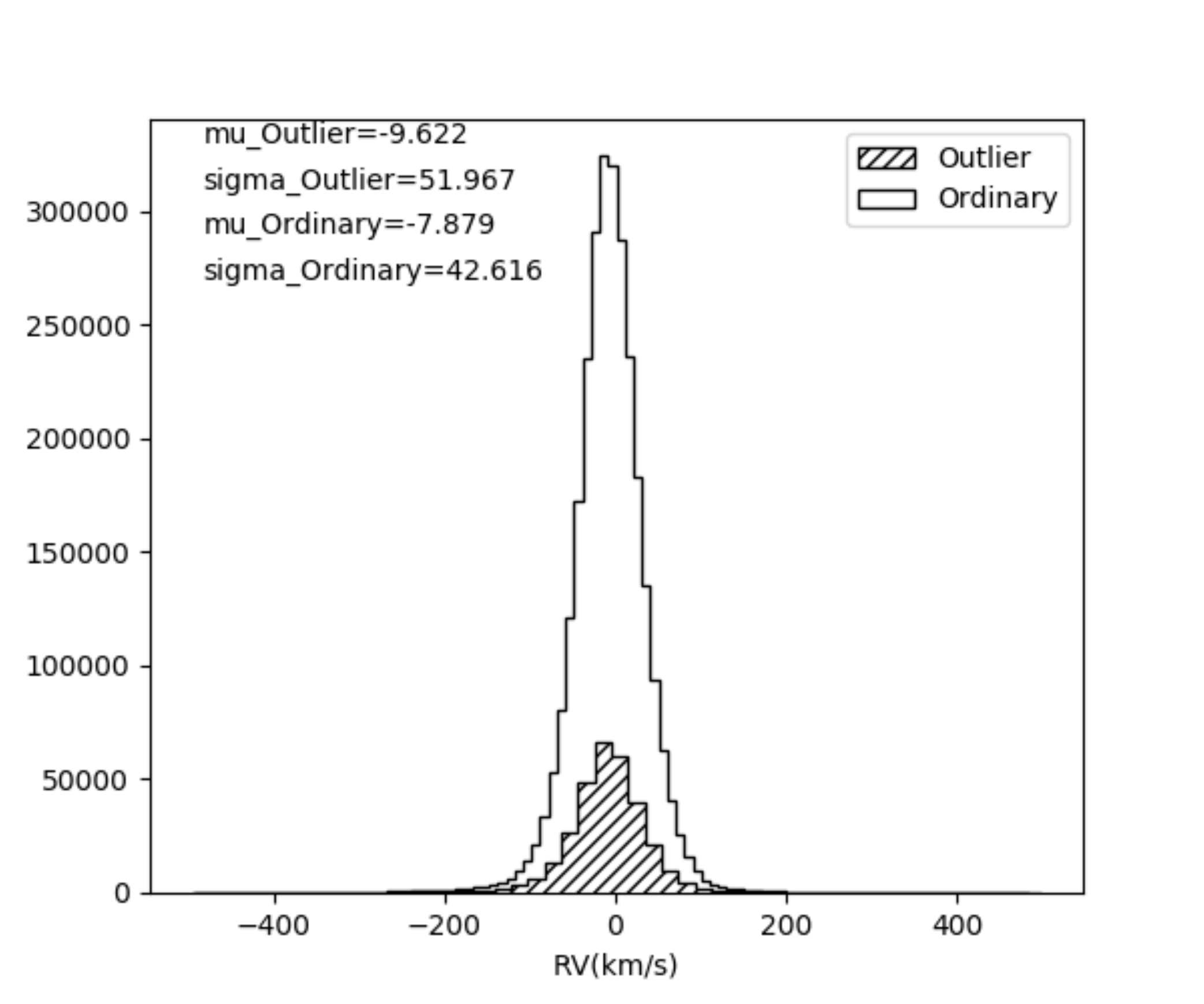}
	\caption{ The comparison of parameter RV released from LAMOST DR5 for outliers and the Ordinary, and the numbers of bins are 50 and 100 respectively. We can find the standard deviation of outliers' RV is a little larger than that of the rest. }
	\label{Fig11}
\end{figure}

In order to test the validity of the selected outlier spectra in dataset $OD$, we compare the properties of outlier spectra with those of the ordinary spectra. First we compare the basic atmospheric physical parameters (effective temperature, surface gravity and metallicity) of outliers to the ordinary spectra, which is shown in $Figure~\ref{Fig9}$. Second is the comparison of S/N, and from Figure~\ref{Fig10} we can see that the S/N of outliers are much lower than the ordinary ones in a whole, and the mean values of S/N for outliers and the ordinary are 64.1 and 91.8 respectively. Next the comparison of RV between outliers and the ordinary is shown in Figure~\ref{Fig11}. Under the same calculation accuracy of LAMOST pipeline, we can find the standard deviation of outliers' RV is a little larger than that of the ordinary. Through the analysis of the nature of the outlier spectra, we found that the outlier spectra found using our method are validate.
\subsection{LOF cut for highlight particularity}
\label{sect:LOF cut}

We have proved the validity of the outlier spectra we pick out, and that ensures we can select more representative special spectra to study the relationship between outliers and parameters. So, in order to get a sub sample with the most highlight particularity, we make the LOF cut process using a criterion applied to our outlier dataset $OD$ using the cumulative distribution function (CDF for short). In detail, we get a new cut criterion by making a new threshold from second derivative of the CDF curve. 

In probability theory and statistics, the CDF of a real-valued random variable $X$, is the probability that $X$ will take a value less than or equal to $x$, evaluated at $x$. If we consider the LOF values of outlier as a continuous random variable, its cumulative distribution curve is monotonically increasing. With the increase of LOF values, the CDF value tends to 1 (100$\%$), thus we use a polynomial function to fit the curve and the first derivative and the second derivative can be used to obtain the intercept point. We calculate the second derivative and find a point for the LOF values, that the mathematical symbols of the second derivative of the points in front of it is different from those of the points after it. This shows that the slope of the function curve at this point (the first derivative) has the maximum rate of change. So we take this point as the inflection point of the CDF function, that is, our intercept point. The intercept point locates at LOF value of 3.01939, which gives us 3,627 most outlier spectra as our special sub sample. In the following part we give a new name to those 3,627 most outlier spectra, sample $MO$, to make further research. 
\subsubsection {Parameters comparison with APOGEE}
\label{param com}

\begin{figure}
	\centering
	\includegraphics[width=1\textwidth, angle=0]{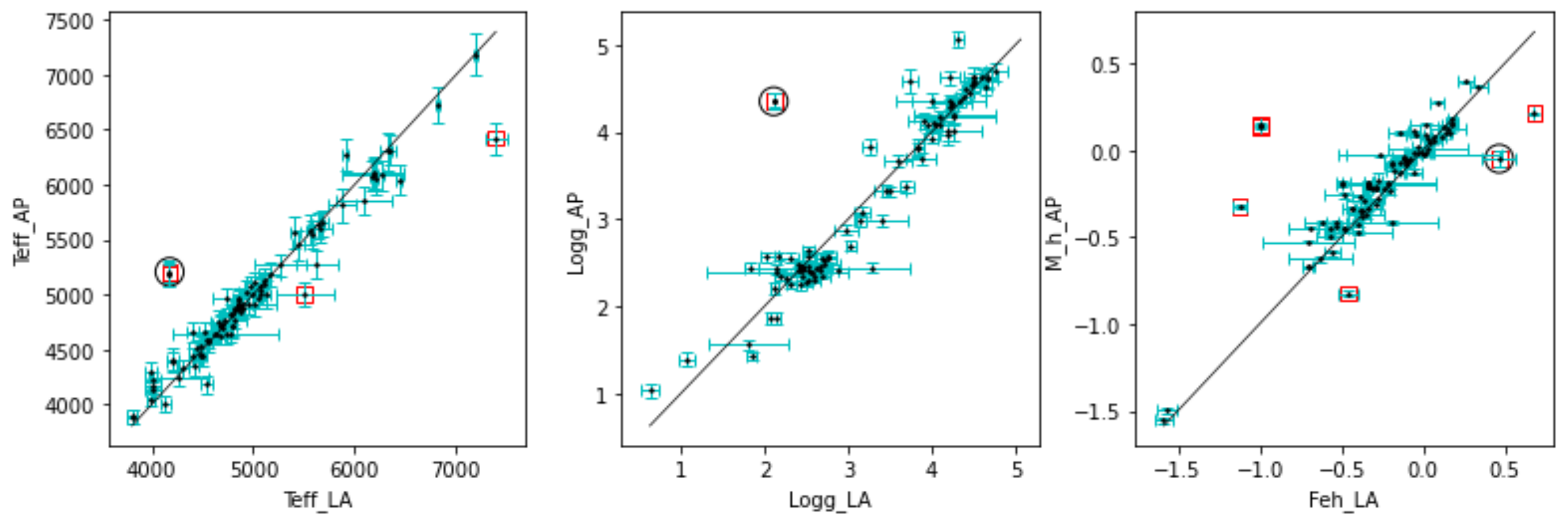}
	\caption{ The comparison between released parameters generated from LASP and the parameters from APOGEE for the common spectra in $MO$. The solid oblique line represents 1:1. }
	\label{Fig12}
\end{figure}

It is generally believed that the higher the spectral resolution, the higher the accuracy of the parameters. We decided to cross-compare the outlier spectrum with APOGEE to study the properties of the outlier spectrum. As J\"{o}nsson pointed out that the Apache Point Observatory Galactic Evolution Experiment (APOGEE) was originally an infrared stellar spectroscopic survey within Sloat Digital Sky Survey (SDSS)-III (\cite{2020AJ....160..120J}; \cite{2017AJ....154...94M}; \cite{2011AJ....142...72E}). APOGEE DR16 includes about 430,000 spectra for stars having high-resolution (R~22,500) , near-infrared (1.514-1.694$\mu$m), and it covers both the northern and the southern sky providing stellar parameters and chemical abundances of up to 26 species \cite{2020AJ....160..120J}. Using a maximum radius of 3 arcsec, we cross the sample $MO$ with APOGEE and get 122 common spectra to analysis the three basic physical parameters: $T$ $_{eff}$, log $g$, and $[$Fe$/$H$]$. In our work each spectrum is treated as an independent observation, and we show the comparison of the released parameters of these common spectra in Figure $\ref{Fig12}$. The left, centre and right sub figures show the $T$ $_{eff}$ plane, the log $g$ plane and the $[$Fe$/$H$]$ plane respectively. The parameters from LASP are shown along the $x$-axis and the parameters from APOGEE along the $y$-axis, with cyan error bar representing the uncertainty of parameters. The results of the three experiments show that the parameters of our outlier spectra obtained by LASP are in agreement with APOGEE on the whole, except for a few which deviate far from the 1:1 line. We further analyse the spectra with deviation of greater than 2 times of variance for each parameter that are marked with red boxes in the figure. We find that the larger parameter deviation is due to the bad points in the spectrum, which may cause errors in the parameter estimation of LASP (LASP derive the stellar parameters via minimizing the squared difference between the observations and the model in the range 4400-6800$\mathring{A}$, which is illustrated in \cite{b4}). For example, the spectrum marked with a gray circle in the three sub figures has the largest deviation in all three parameter, and this spectrum (LAMOST J030136.37-000922.6) has abnormal flux values in the range 4700-5200$\mathring{A}$. For details of the comprehensive comparison between APOGEE and LAMOST, one can refer to \cite{2018A&A...620A..76A}. 
\subsubsection {Clustering for the outliers sample $MO$}
\label{sect:cluster}
In order to better explain the properties of the most outlier spectra, we divide the sample $MO$ into 10 clusters and carry out spectral analysis for each cluster. K-means clustering is used in this part, and it is a method of clustering analysis that aims to divide all samples into $k$ clusters and each sample belong to one single cluster, objects in the same cluster being similar to each other. The number of clusters needs to be specified before operating the traditional method K-means. For the purpose of analysing the characteristics of the most outlier spectra with different outlier morphology and specific properties, firstly we apply the classical K-means clustering method and appoint the number of clusters $k$=50 to the sample $MO$. According to the prior knowledge of the continuous spectrum and spectral lines of the spectra, by defining the number of 50 clusters, spectra in sample $MO$ can be divided into sufficient clusters based on the details of spectral differences. Some spectral fluxes may suddenly change to zero or invalid value on different wavelength range. Another phenomenon can form long or short emission lines, and the span may be only a few angstroms or even reach more than half the wavelength range of the whole spectrum. Only when the spectra are divided into sufficient number of clusters, can we combine these differences and analyse them comprehensively according to the prior knowledge of astronomical spectra. After merging similar clusters manually we classify sample $MO$ into 10 groups finally. For example, spectra that contain a small number of bad points of flux values are grouped together, no matter which wavelength range these points are in. Detailed description of those groups are shown as below:

\begin{enumerate}
	
	\item[(1)] Group 1 of flux having many bad points. 
	
	The outlier spectra in this group have such characteristics that the fluxes at some wavelength points suddenly become zeros or abnormal values. These zeros or abnormal values can be renamed as bad points in the flux of a spectrum, and the wave band length of the part with bad points of the flux is more than 200 angstroms for every spectrum in this group. There are 203 spectra in this group, and according to general coding habits the number starts from 0 in our catalog.
		
	\item[(2)] Group 2 of flux having fewer bad points.
	
	Being similar to Group1, the fluxes at some wavelength points suddenly becomes zeros or abnormal values for spectrum in this group. But the wave band length of the part with bad points of the flux is less than 200 angstroms for every spectrum in this group. There are 426 spectra in Group 2.
	 
	\item[(3)] Group 3 of flux having false emission lines.
	
	In this group the fluxes of a spectrum suddenly become very strong at certain wave points. At first glance, the strong fluxes look like emission lines. Only by careful observation can we find that these sudden strong flow points are abnormal values in a flux, and we also call them bad points. The number of items in this group is 963.

	\item[(4)] Group 4 of flux having anomalies caused by stray light pollution.
	
	There are 823 spectra in sample $MO$ that the continuum of spectrum is slightly convex like a small hill at roughly wavelength range 5400-5500$\mathring{A}$, and they are classified into Group 4. We carefully check out these spectra and find that all of them are observation output of the telescope on just the same day. This phenomenon should be caused by stray light pollution in the sky.
	
	\item[(5)] Group 5 of flux showing nebular contamination.
	
	Group 5 includes 189 spectra totally and all of them show strong nebular characteristics, including stellar formation region (HII), supernova remnant (SNR), and planet nebular (PNe). We try to spectroscopically classify them with an emission-line diagnostic criterion based on 
	
	$ log([SII] \lambda6718, 6732/H\alpha) $ and $ log([NII] \lambda6585/H\alpha) $, Using the following equation:
	$ log(\frac{[SII]}{H\alpha}) \ge 0.63log(\frac{[NII]}{H\alpha}) -0.55$ (\cite{2002RMxAC..12..174R}; \cite{2003A&A...400..511M}; \cite{2008MNRAS.384.1045K}). Figure~\ref{Fig13} shows the classification of these spectra having nebular characteristics got using their spectral line ratios. 
	
	\begin{figure}
		\centering
		\includegraphics[width=0.4\textwidth, angle=0]{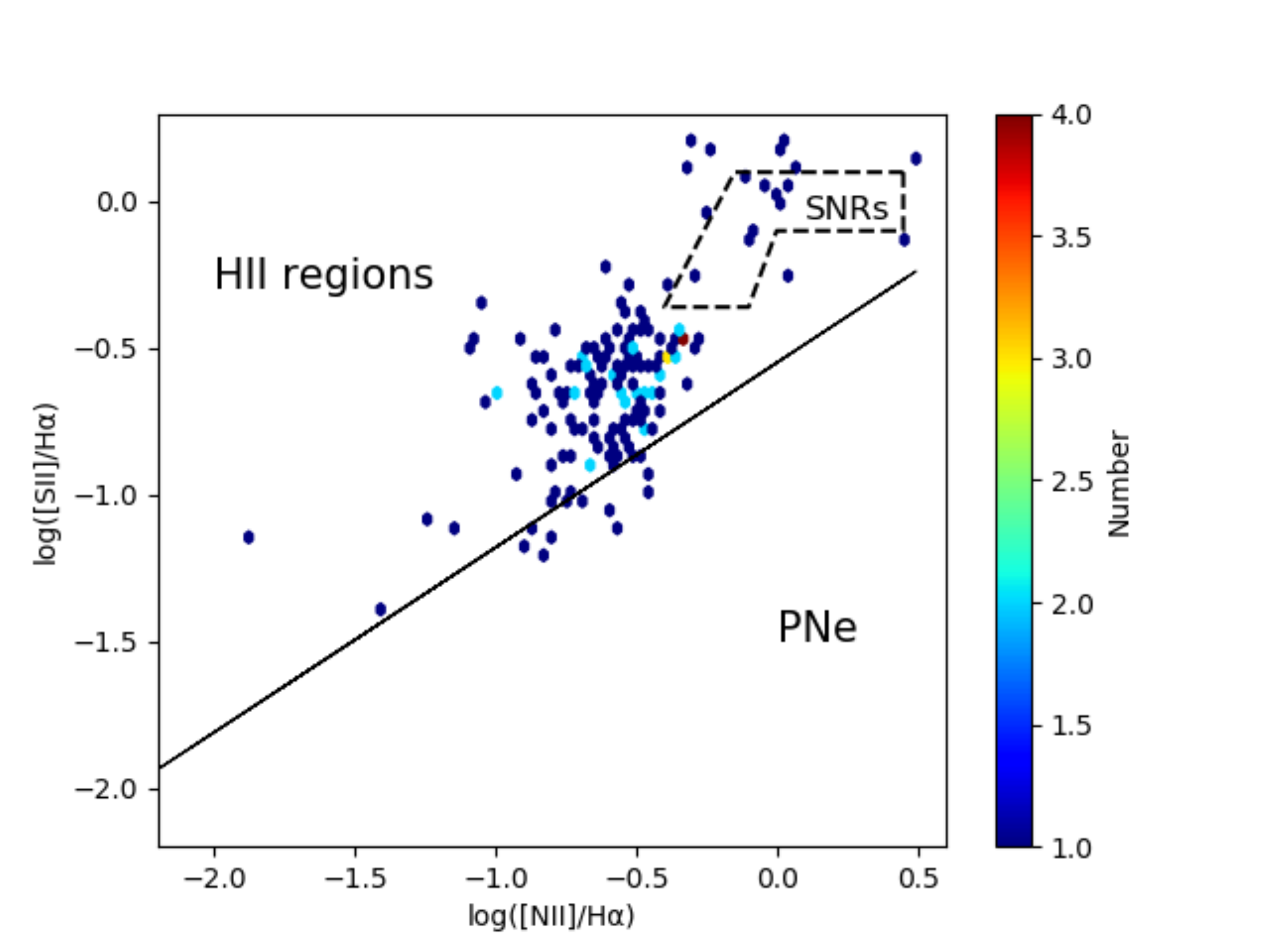}
		\caption{ Spectra in Group 5 show strong nebular characteristics of HII, SNR or PN, and all the spectra are classified according to their spectral line ratio. The solid line represents the separator of HII/PNe. }
		\label{Fig13}
	\end{figure} 
	
	\item[(6)] Group 6 of flux having bad points in the region of wrongly connection between blue and red bands.
	
	For the spectra in this group there are bad flux values in the region of wrongly connection between blue and red bands. And Group 6 contains 669 spectra.
	
	\item[(7)] Group 7 of normal spectra.
	
	This group include 44 normal spectra such as 13 A type stars, and this is explicable. Our analysis method is applied to the dataset of archived AFGK stars in LAMOST DR5, and the A stars in the dataset only include late types which occupy 1.73\% of the total dataset. Thus late A type stars may be picked out as outliers.    
	
	\item[(8)] Group 8 of spectra having continuum anomalies.
	
	The red band of 38 spectra in this group are seriously reddening, and there are 78.95\% of  the stars corresponding to them are located within $\pm30$ degrees of the Galactic latitude of the celestial coordinate system (according to \cite{2016ApJS..227...27D}, there are variations of reddening between different stars especially at low Galactic latitudes). For the other 21 spectra, the shape of continuous spectrum is abnormal may due to bad flux calibrations. Perhaps some spectrum of peculiar stars may be found in this group.
	
	\item[(9)] Group 9 of flux having bad points in the red band only.
	
	The spectra in this group exhibit similar characteristics, that is, the flux in the blue band of the spectrum is normal value, while the flux in the red band is obviously abnormal. There are 118 spectra in this group.
	
	\item[(10)] Group 10 of flux with bad points in the blue band only.
	
	The red band of the spectrum in this group is normal, while the blue band has obvious continuum anomalies. And the amount of spectra in this group is 133.
\end{enumerate}
\subsubsection{Distribution of $MO$ in LAMOST empirical library parameter space}

\begin{figure}
	\centering
	\includegraphics[width=1\textwidth, angle=0]{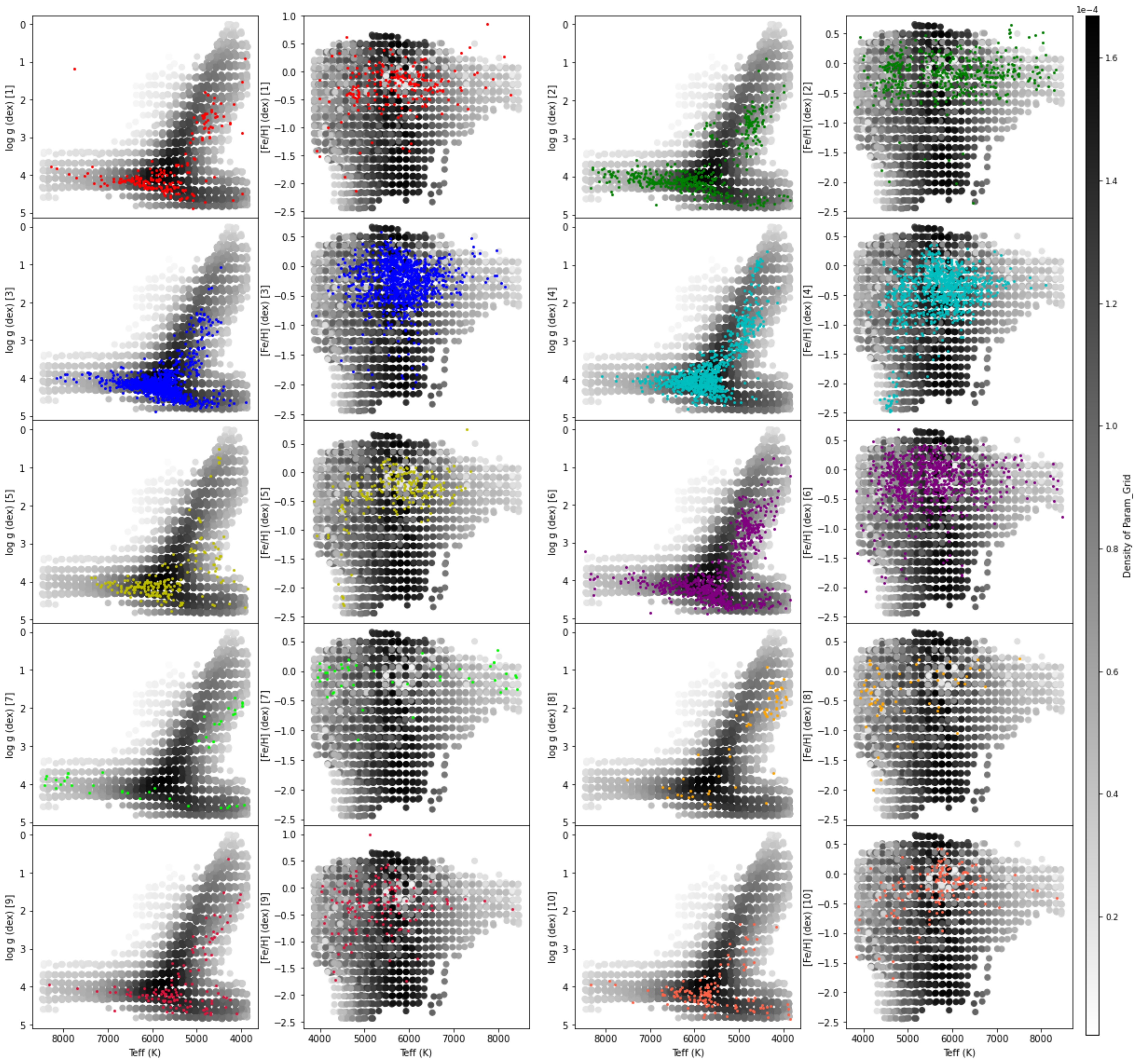}
	\caption{ The distribution of published parameters of 10 groups of $MO$ in the parameter space of the LAMOST DR5 empirical library, each row showing two groups and the name of the group is marked at the end of the $y$-axis title of each sub picture, enclosed in square brackets. The background of the gray gradient represents the density distribution of the template parameters, and the points of different foreground colours indicate the parameter distribution of our outlier spectra. }
	\label{Fig14}
\end{figure}

Du et al.(2019) presented an empirical stellar spectra library using spectra of LAMOST DR5, containing a uniform dataset in which $T$ $_{eff}$ range from 3750 to 8500 K , log $g$ from 0 to 5.0 dex, and $[$Fe$/$H$]$ from -2.5 to +1.0 dex \cite{2019ApJS..240...10D}. The full wavelength coverage of this empirical library is from 3800-8900$\mathring{A}$. Figure $\ref{Fig14}$ shows the distribution of published parameters of our most outlier spectra in $MO$ in the parameter space of the empirical library. We show the parameters according to our grouping of $MO$ explained in $\ref{sect:cluster}$, each row showing two groups and each group having the same foreground colour. For each group the left and right sub figures show the $T$ $_{eff}$-log $g$ plane and the $T$ $_{eff}$-$[$Fe$/$H$]$ plane respectively, and the name of the group is marked at the end of the $y$-axis title of each sub picture, enclosed in square brackets. The background of the gray gradient represents the density distribution of the library parameters, and the points of different foreground colour indicate the parameter distribution of our outlier spectra. Most of the published parameters of our outlier spectra fall within the parameter grid of the library spectra, and the density distribution also conforms to the distribution trend of the library spectra. Group 3, Group 7, Group 8 and Group 9 are more consistent examples. The spectral false emission lines in Group 3 are masked out when processed by LASP, so the parameter results are good. Group 7 contains normal spectra so they make good agree. From Group 8 we can find that spectral continuum anomalies such as reddening have little effect on LASP. LASP mainly relies on the blue band for parameter estimation as explained in $\ref{param com}$, so the red band anomaly has less impact, that's the fact of Group 9. While Group 1, Group 2, Group 4 and Group 10 have a low degree of conformity with the parameter grid. Group 1 has the lowest degree of coincidence. The point at the top left corner of the left sub picture and the point at the top right corner of the right sub picture of Group 1 represents the same spectrum of LAMOST J065309.31+252943.7, and the abnormal flux values in 5000-5700$\mathring{A}$ of the spectrum lead to large deviation of the estimated parameters. The band range occupied by bad points in Group 2 is smaller than that in Group 1, which also affects the parameter estimation. The convexities at the blue band of the spectrum in Group 4 may cause some metal poor misjudgements, as shown in the lower left corner of the right sub picture. The blue band has obvious continuum anomalies in some bands of Group 10, so the parameter estimation are also affected.

\section {Conclusion}
\label{sect:conclusion}

In this work, we download 5,349,401 spectra in LAMOST DR5 which have released parameters such as temperature, gravity and metallicity, and we only select spectra that containing more than $\frac{2}{3}$ good quality sampling points which remain 3,134,236 spectra. Next we use the PCA method to reduce the dimension of these spectra and their outlier factors are calculated by LOF method for outlier detection. According to the outlier factors we first pick out 313,424 (the top 10 percent of 3,134,236) outlier spectra to find that the parameters of outlier spectra have acceptable deviation compared with the ordinary spectra. Second we get 3,627 most outlier spectra (dataset $MO$) using a new cut on the outlier factors (threshold of LOF value 3.01939) to make further research. We cross $MO$ with APOGEE getting 122 common spectra and after comparison we get the conclusion that the parameters of our most outlier spectra obtained by LASP are in agreement with APOGEE on the whole. Meanwhile, we also find that the larger parameter deviation is due to the bad points in the spectrum, which may cause errors in the parameter estimation of LASP. Then we classify all spectra in $MO$ into 10 groups using method based on K-means and give detail description of those groups. We also find that most of the published parameters of $MO$ fall within the parameter grid of the LAMOST DR5 empirical library, and the density distribution also conforms to the distribution trend.

To sum up, the result of our work shows that most of the stellar parameters are mostly corrected according to external comparison even the spectra have lots of bad pixels or bad flux calibrations, which means the LASP has good adaptability. On the other hand, some outlier spectra show strong nebular contamination and the corresponding parameters should be carefully used. We provide a catalog and an atlas in electronic form in the online version of all the 3,627 outlier spectra. The catalogue contains coordinates, names and group information that shown in Table \ref{table1}, and the atlas includes pictures of them.

\begin{table}[htbp]\footnotesize
	\centering
	\caption{catalog of 3627 outlier spectra}
	\begin{tabular}{ccccc}
		\hline
		RA & DEC & FILENAME & GROUP &  \\ \hline
		292.17819 & 41.149151 & spec-57307-KP192102N424113V03\_sp07-153.fits & 2 &  \\ \hline
		57.855423 & 1.550066 & spec-56219-EG035637N030328B01\_sp02-038.fits & 2 &  \\ \hline
	\end{tabular}
	\label{table1}
\end{table}

\section *{Acknowledgment}

This work is supported by the National Natural Science Foundation of China (NSFC) under Nos.U1931209, 11903008, U1931106, and the Science Foundation of DeZhou University (grant No.2019xjrc39). Guoshoujing Telescope (the Large Sky Area Multi-Object Fiber Spectroscopic Telescope, LAMOST) is a National Major Scientific Project built by the Chinese Academy of Sciences. Funding for the project has been provided by the National Development and Reform Commission. LAMOST is operated and managed by the National Astronomical Observatories, Chinese Academy of Sciences.

\section*{References}

\bibliography{work1}

\end{document}